\documentclass[useAMS,usenatbib]{mnras}
\usepackage{times}

\usepackage{graphicx}	% Including figure files
\usepackage{amsmath}	% Advanced maths commands
\usepackage{amssymb}	% Extra maths symbols
\usepackage[usenames, dvipsnames]{color}
\usepackage{indentfirst}

\newcommand {\bc}{\begin {center}}
\newcommand {\ec}{\end {center}}
\newcommand {\be}{\begin {equation}}
\newcommand {\ee}{\end {equation}}
\newcommand {\beq}{\begin {eqnarray}}
\newcommand {\eeq}{\end {eqnarray}}

\newcommand {\ergs}{{\rm erg\ \rm s^{-1}}}

\newcommand {\comment}[1]{}

\def\lbar {\lambda\hskip-5pt\raise3pt\hbox {--}}
\def\lbr {\lambda\raise2pt\hbox {\hskip-4pt{\scriptsize --}}_\C}

\renewcommand{\d}{{\rm d}}

\renewcommand{\d}{{\rm d}}

\title[Aperiodic variability in X-ray pulsars]
{{Broadband} aperiodic variability in X-ray pulsars: accretion rate fluctuations propagating under {the} influence of viscous diffusion}
\author[A. A.~Mushtukov et al.] 
{Alexander~A.~Mushtukov,$^{1,2,3}$\thanks{E-mail: al.mushtukov@gmail.com (AAM)} 
Galina~V.~Lipunova,$^{4}$ 
Adam Ingram,$^{5}$ 
\newauthor
Sergey S. Tsygankov,$^{6,3}$
Juhani M\"onkk\"onen,$^{6}$
Michiel van der Klis$^{2}$\\ 
$^1$ Leiden Observatory, Leiden University, NL-2300RA Leiden, The Netherlands \\
$^2$ Anton Pannekoek Institute, University of Amsterdam, Science Park 904, 1098 XH Amsterdam, The Netherlands \\
$^3$ Space Research Institute of the Russian Academy of Sciences, Profsoyuznaya Str. 84/32, Moscow 117997, Russia \\
$^4$ Sternberg Astronomical Institute, Moscow Lomonosov State University, Universitetski pr. 13, Moscow 119234, Russia \\
$^5$ Department of Physics, Astrophysics, University of Oxford, Denys Wilkinson Building, Keble Road, Oxford OX1 3RH, UK \\
$^6$ Department of Physics and Astronomy, University of Turku, FI-20014 Turku, Finland \\
} 
\pubyear{2019}

\begin{document}
\label{firstpage}
\pagerange{\pageref{firstpage}--\pageref{lastpage}}
\maketitle

%%%%%%%%%%%%%%%%%%%%%%%%%%%%%%%%%%%%%%%%%%%%%%%%%%%%%%%%%%%%%%%%%%%%%%%%%%%%%%
%% Abstract, Keywords and contact details                                   %%
%%%%%%%%%%%%%%%%%%%%%%%%%%%%%%%%%%%%%%%%%%%%%%%%%%%%%%%%%%%%%%%%%%%%%%%%%%%%%%
\begin{abstract}
\indent
{
We investigate aperiodic X-ray flux variability in accreting highly magnetized neutron stars - X-ray pulsars (XRPs).
The X-ray variability is largely determined by mass accretion rate fluctuations {at the} NS surface, which replicate accretion rate fluctuations at the inner radius of the accretion disc.
The variability at the inner radius is due to fluctuations arising all over the disc and propagating inwards under the influence of viscous diffusion. 
The inner radius varies with {mean} mass accretion rate and can be estimated from {the} known magnetic field strength and accretion luminosity of XRP{s}. 
Observations of transient XRPs covering several orders of magnitude in luminosity give a unique opportunity to study effects arising due to the changes of the inner disc radius. 
We investigate {the} process of viscous diffusion {in XRP accretion discs}
and construct new analytical solutions of the diffusion equation applicable for thin accretion discs truncated both from inside and outside.
Our solutions are the most general ones derived in the approximation of Newtonian mechanics.
We argue that the break observed at high frequencies in the power density spectra of XRPs corresponds to the minimal time scale of the dynamo process, which is responsible for the initial fluctuations.
Comparing data {from the} bright X-ray transient A~0535+26 with our model, we conclude that the time scale of initial variability in {the} accretion disc is a few times longer than {the} local Keplerian time scale.
}
\end{abstract}


\section{Introduction}
%\green{\textit{[AI: I've just realized, what do you assume for $\Delta R(R)/R$? I guess you assume that it is constant with radius? We should say something about this somewhere. I guess for $\Delta R(R) / R =$constant (which is more than reasonable of course), there is a degeneracy between $F_{var}$ and $\Delta R(R) / R =$, so I guess technically the model parameter is $(F_{var} \Delta R / R)$. We could say something to this effect in Section 4.2 maybe, just a few sentences.]}}
X-ray pulsars (XRPs) are highly magnetized (typical magnetic field strength at the surface $\gtrsim 10^{12}\,{\rm G}$) neutron stars (NSs) in close binary systems, whose luminosity in X-rays is caused by accretion from {the} companion star \citep{2015A&ARv..23....2W}. 
The accretion flow, in {the} form of {a} stellar wind or accretion disc, is interrupted by {the} strong {NS} magnetic field ($B$-field) {at the \textit{magnetospheric radius}, $R_{\rm m}$. This}  
%at a certain distance (called magnetospheric radius) from the NS. 
%The magnetospheric radius $R_{\rm m}$ 
radius depends on the accretion flow geometry, mass accretion rate $\dot{M}$ and $B$-field strength and structure \citep{1972A&A....21....1P,1973ApJ...179..585D,1978SvA....22..702L,1980A&A....86..192A}. 
If the mass accretion rate is sufficiently high {for matter to} go through the centrifugal barrier caused by rotation of {the} strongly magnetized NS, the accretion flow penetrates into the magnetosphere and follows magnetic field lines to reach the NS surface in regions located close to the magnetic poles. 
There the accretion flow loses its kinetic energy, which is radiated mostly in {the} X-ray energy band. 
Misalignment between {the} rotational and magnetic axes results in the phenomenon of XRPs. 
If the matter cannot penetrate through the centrifugal barrier, {the} accretion process stops, leading to {the} so-called "propeller" effect \citep{1975A&A....39..185I,1977SvAL....3..138S,1985A&A...151..361W,2004ApJ...616L.151R,2016MNRAS.457.1101T,2016A&A...593A..16T}.

{The} accretion luminosity of known XRPs covers several orders of magnitude from $\sim 10^{33}\,\ergs$ up to $\sim 10^{40}\,\ergs$.
Many sources are transients and show significant variability of luminosity. 
 The brightest XRPs belong to the recently discovered class of pulsating ultra-luminous X-ray sources (ULXs, see e.g. \citealt{2014Natur.514..202B,2017Sci...355..817I}). The observed X-ray luminosity in pulsating ULXs exceeds the Eddington value, which is $L_{\rm Edd}\approx 2\times 10^{38}\,\ergs$ for {a} NS, by a factor of hundreds \citep{2017Sci...355..817I}.

\begin{figure}
\centering 
\includegraphics[width=7.cm]{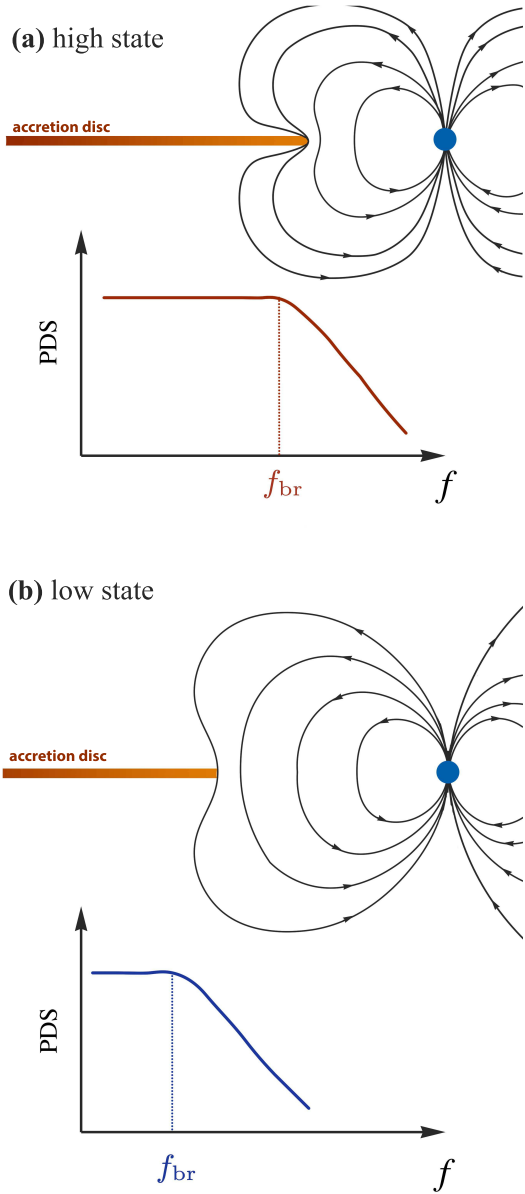}
\caption{ 
The broadband aperiodic variability of the accretion luminosity in XRPs is originated from the variability of the mass accretion rate, which is initially produced in the disc. 
The disc in XRP is truncated due to interaction with a strong magnetic field of a NS (see \citealt{1987ans..book.....L} and \citealt{2014EPJWC..6401001L} for review). 
The higher frequencies of the variability are introduced to the accretion flow at smaller radial coordinates.
Because the inner disc radius decreases with the increase of the mass accretion rate, the XRPs in high luminosity state (a) can produce aperiodic variability at higher Fourier frequency in respect to XRPs at low luminosity state (b). 
In particular, the break in PDS associated with the inner disc radius shifts towards higher Fourier frequencies.
} 
\label{pic00}
\end{figure}

The X-ray flux originates from regions located close to the NS magnetic poles. 
The geometry of the radiating regions can be affected by radiation pressure and is determined by the mass accretion rate. 
At relatively low mass accretion rates ($\lesssim 10^{17}\,{\rm g\, s^{-1}}$), the accretion flow is stopped by Coulomb collisions and/or plasma oscillations at surface layers of {the} NS \citep{1969SvA....13..175Z}, while at high mass accretion rates ($\gtrsim 10^{17}\,{\rm g\, s^{-1}}$), the radiation pressure becomes strong enough to stop the flow at some height above the stellar surface \citep{1976MNRAS.175..395B,2015MNRAS.447.1847M}. In the latter case, {an} accretion column forms. The column can provide luminosity well above the Eddington value because it is confined by {the} strong magnetic field, {and thus} radiation pressure can be largely reduced due to the reduction of scattering cross-sections in a strong B-field \citep{1981A&A....93..255W,2015MNRAS.454.2539M}, {plus} photon bubbles can come into play reducing effective radiation pressure \citep{1992ApJ...388..561A}. 
The radiation of the accretion column is likely beamed towards the NS surface \citep{1976SvA....20..436K,1988SvAL...14..390L} and, thus, a large fraction of X-ray flux is reprocessed/reflected by the atmosphere \citep{2013ApJ...777..115P}.

XRPs show strong aperiodic variability of X-ray flux over a very broad frequency range similar {(modulo mass scaling)} to what is detected in accreting black holes (BHs, see e.g. \citealt{2000A&A...363.1013R}) and active galactic nuclei (AGN, see e.g. \citealt{2004MNRAS.348..783M}). 
The time scale of the observed variability extends down to milliseconds. 
Comparing accreting NSs with BHs we note that the power density spectra of weakly magnetized NSs contains much stronger variability at frequencies close of one kHz \citep{2000A&A...358..617S}.
The power density spectrum (PDS) typically includes a broad component,
which can be approximated by {a} broken (or double broken) power-law \citep{1993ApJ...411L..79H}, and narrow features that {are} classified as quasi-periodic oscillations (QPOs, see \citealt{1994ApJ...436..871T}). Both components are detected to {evolve} with long term trends in the observed luminosity (\citealt{2009A&A...507.1211R}).

The aperiodic variability in X-ray binaries and AGNs is naturally explained by the propagating fluctuations model \citep{1997MNRAS.292..679L,2001MNRAS.321..759C,2007ApJ...660..556T}. According to the model, the initial fluctuations of the mass accretion rate arise at different radial coordinates in {the} accretion disc and then propagate inwards and outwards, modulating the fluctuations arising at other radial coordinates. In this scenario, different time scales are injected into the accretion flow at different distances from the central object, while the observed variability of X-ray flux reflects the variability of the mass accretion rate at the inner parts of accretion disc \citep{2001MNRAS.327..799K, 2013MNRAS.434.1476I, 2016AN....337..385I,2018MNRAS.474.2259M}. 

There is a fundamental difference between the X-ray flux variability in accreting BHs and XRPs: in the case of BH binaries, an observer detects X-ray photons originating from the accretion disc itself, while in the case of XRPs, photons (seed photons at least, see below) mostly originate from the NS surface. 
We can imagine a few sources of aperiodic variability in XRPs:
(i) the variability of X-ray flux due to the mass accretion rate fluctuations at the inner disc radius, which determines directly the variability of the mass accretion rate at the NS surface;
(ii) the variability caused by changes of {the} beam pattern due to changes {in the} geometry of {the} radiating region \citep{2018MNRAS.474.5425M}; 
(iii) the variability due to reprocessing of seed X-ray photons by the accretion flow (in {the} form of {a} hot optically thin accretion flow at low mass accretion rates, see e.g. \citealt{1995ApJ...452..710N}, or {an} optically thick accretion flow at high mass accretion rates, see e.g.  \citealt{2017MNRAS.467.1202M});
(iv) production of photon bubbles in {the} accretion column at high mass accretion rates \citep{1992ApJ...388..561A,2006ApJ...643.1065B}.
The time scales of photon emission and diffusion in radiating regions {also} affect {the} observed variability properties, but {only on} time scales much {shorter} than {those} of {the} processes discussed above.
{In this paper, we focus on the aperiodic variability due to propagating fluctuations of the mass accretion rate in a disc, and assume that the fluctuations in X-rays replicate fluctuations of the mass accretion rate at the inner disc radius.}

The changes of PDS with luminosity of the XRP are largely caused by {changes} of the inner disc radius, which depends on the mass accretion rate (see Fig.\ref{pic00}, see e.g. \citealt{2009A&A...507.1211R}).
In a number of cases, the magnetic field strength of the XRP is known (it can be measured from cyclotron line scattering features in the X-ray spectrum), and thus the disc inner radius $R_{\rm m}$, and the Keplerian frequency there, can be estimated. The similar behavior of the observed break frequency and estimated Keplerian frequency at the magnetospheric radius reveals that the changes in PDS break frequency with X-ray luminosity are largely caused by changes in $R_{\rm m}$ \citep{2009A&A...507.1211R}. The power-law dependence of break frequency on mass accretion rate is consistent with the break frequency being proportional to the Keplerian frequency at $R_{\rm m}$. Constraint of the constant of proportionality obtained through detailed physical modelling offers the opportunity to use the measured PDS break frequency to estimate $R_{\rm m}$ for XRPs with poorly constrained magnetic field strength. Moreover, such constraints will feed into our understanding of accretion disc variability in general, providing diagnostics of the accretion flow geometry of XRBs and AGN through timing properties. Similarly, the low frequency breaks detected in some XRPs contain information about the outer disc radius \citep{2005astro.ph..1215G}.

In this paper, we focus on XRPs where accretion {takes place} through {a} geometrically thin accretion disc (this assumption puts limitations on {the} range of mass accretion rate under consideration). 
We focus on variability caused by processes in {the} accretion disc: mass accretion rate fluctuations arising and propagating in {the} disc due to the process of viscous diffusion.
We analyze the effects arising {from} truncation of {the} accretion disc at certain inner $R_{\rm in}$ and outer $R_{\rm out}$ radii,
assuming that the disc loses its mass through the inner radius only.
We consider the general case of non-zero torque at the inner disc radius and provide a theoretical background for calculations of the PDS in accreting strongly magnetized NSs.

{
The paper consists of five sections. 
In Section \ref{sec:AD_basic_conditions} we discuss the basic features of accretion discs in XRPs at different luminosity states, accretion disc geometry and typical time scales in the accretion flow.
The analytical theory of propagating fluctuations of the mass accretion rate is discussed in Section \ref{sec:PropFluxTheory}, where we introduce a new analytical solution of the viscous diffusion equation accounting for disruption of {the} accretion disc both from {the} inside and outside
(see Section \ref{sec:GfLipunova}). 
Section \ref{sec:NumResults} presents our numerical results based on the theory and novel Green functions developed in Section \ref{sec:PropFluxTheory}.
Summary and conclusions are given in Section \ref{sec:Summary}.
}

\section{Accretion discs in X-ray pulsars}
\label{sec:AD_basic_conditions}

In this paper, we will define and apply a model for the aperiodic X-ray variability of X-ray pulsars. In this Section we first explore the expected disc geometry (Section \ref{sec:ADGeometry}) and accretion regime (Section \ref{sec:AccRegime}) before considering typical timescales of the system (Section \ref{sec:TimeScale}), and finally commenting on specific features of the accretion flow in some classes of XRPs that we do not consider here, but must be included in advanced models of aperiodic variability (Section \ref{sec:AD_in_classes})

\subsection{Accretion disc geometry}
\label{sec:ADGeometry}

{The} accretion disc in XRPs is truncated at the magnetospheric radius, $R_{\rm m}$, which can be roughly estimated {by comparing} the $B$-field pressure {with the} ram pressure of {the} accreting material \citep{1987ans..book.....L,2002apa..book.....F}:
\be\label{eq:Rm}
R_{\rm m}=2.4\times 10^{8}\Lambda B_{12}^{4/7}L_{37}^{-2/7}m^{1/7}R_{\rm NS,6}^{10/7} \,\,\,{\rm cm},
\ee
where $\Lambda<1$ is a constant which depends on the accretion flow geometry, with $\Lambda=0.5$ being a commonly used value for the case of accretion through the disc \citep{1979ApJ...232..259G,2014EPJWC..6401001L}, $B_{12}$ is {the} magnetic field strength $B$ at the NS surface in units of $10^{12}\,{\rm G}$, $L_{37}$ is the accretion luminosity $L$ in units of $10^{37}\,\ergs$, $m$ is {the} NS mass in units {of} solar masses $M_\odot$, and $R_{\rm NS,6}$ is the NS radius in units of $10^6\,{\rm cm}$. 
Apart from the fact that $\Lambda$ depends on a specific model of the $B$-field and disc interaction, the inner radius of the accretion disc is also affected by the magnetic dipole inclination (see e.g. \citealt{1978SvA....22..702L,1978ApJ...219..617S,1980A&A....86..192A}).
For the typical $B$-field strengths and mass accretion rates in XRPs, the inner disc radius is so large that the disc cannot produce a noticeable fraction of luminosity in {the} X-ray energy band. 
The boundary layer of the accretion disc, where material penetrates into the magnetosphere, is formed due to instabilities developing at the inner disc radius (magnetic Kelvin-Helmholtz and Rayleigh-Taylor, and reconnection, see e.g. \citealt{1990A&A...229..475S,2014EPJWC..6401001L}).
The accretion flow, penetrating into the magnetosphere, follows magnetic field lines and reaches the NS magnetic poles to provide most of the X-ray flux at the surface, where the kinetic energy is converted into heat.

The outer radius of {the} accretion disc is determined by tidal torques in {the} binary system, which removes the angular momentum from the matter near the disc edge and prevents disc spreading. The tidal radius $R_{\rm tid}$ is likely close to the size of the Roche lobe: $R_{\rm tid}\approx (0.8-0.9)R_{\rm L}$ \citep{1977ApJ...216..822P,1977MNRAS.181..441P}.

\subsection{Accretion regime}
\label{sec:AccRegime}

\begin{figure}
\centering 
\includegraphics[width=9.cm]{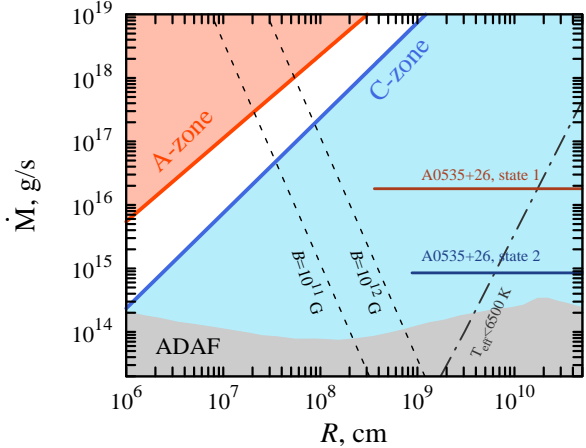}
\caption{
Zones in an accretion disc {for} different mass accretion rates $\dot{M}$.
Inner disc radii for surface magnetic field strengths of $10^{11}$ and $10^{12}\,{\rm G}$ are {represented} by black dashed lines (the magnetic field is assumed to be dipole and the coefficient $\Lambda=0.5$ in equation \ref{eq:Rm}).
The grey area corresponds to the radiatively inefficient zone of optically thin and geometrically thick accretion, where $\sim 90\%$ of {the} accretion flow is in a hot advection dominated state (ADAF, \citealt{1995ApJ...452..710N,2010PASJ...62..661Q}). 
Red and blue zones represent radiation (A-zone) and gas (C-zones) pressure dominated regions in {the} accretion disc \citep{1973A&A....24..337S,2007ARep...51..549S}. 
The black dashed-dotted line represents the radial coordinate where the surface temperature of a thin accretion disc drops below $6500\,{\rm K}$, which results in thermal instability of {the} disc.
Horizontal dark red and dark blue lines represent two states of accretion in {the} transient {source} A~0535+26.
Parameters: $M=1.4M_\odot$, $R=10\,{\rm km}$.
}
\label{pic:DiscZones}
\end{figure}

An optically thick accretion disc can be divided into three zones according to the dominating pressure and opacity sources (see e.g. \citealt{1973A&A....24..337S, 2007ARep...51..549S}). 
Gas pressure and Kramers opacity dominate in the outer regions (C-zone), gas pressure and electron scattering dominate in the intermediate zone (B-zone) and radiation pressure dominates in the inner zone (A-zone). 
Fig.\,\ref{pic:DiscZones} illustrates the boundaries between these three regimes. 
As marked by the dot-dashed line, the effective temperature in the outer parts of the disc may drop below 6500K depending on mass accretion rate. 
In this case, the hydrogen  recombines and a `cooling' wave propagates inward (e.g. \citealt{2001NewAR..45..449L}) and can reach $R_{\rm in}$ making all the hydrogen in the disc neutral. 
Stable accretion from such a "cold disc" state has recently been discovered in a few XRPs  \citep{2017A&A...608A..17T,2017MNRAS.470..126T}. 
The grey region in Fig.\,\ref{pic:DiscZones} denotes {an} advection dominated accretion flow (ADAF) regime \citep{1995ApJ...452..710N}, which is beyond the scope of this paper.

The dashed lines in Fig.\,\ref{pic:DiscZones} show the magnetospheric radius for a selection of typical values of XRP magnetic field strength. 
We see that, for all but the highest accretion rates, the inner disc is expected to be in the C-zone regime. 
Our model for the aperiodic variability XRPs will therefore assume that the disc is entirely in the C-zone. 
We will apply our model to A~0535+26, whose disc extent is also plotted in Fig.\,\ref{pic:DiscZones} for two accretion states. 
The tidal radius of this source is not well known, but that the disc does not appear to be in the cold disc regime implies that $R \lesssim 6\times 10^9$ cm. 
Note that known orbital period and an estimated mass of a companion star in A~0535+26 ($P_{\rm orb}\simeq 110$ days and $M_{\rm comp}\sim 14\,M_{\sun}$, see \citealt{1980A&AS...40..289G}) imply the tidal radius $R_{\rm tid}\gg 6\times 10^9\,{\rm cm}$, but the outer regions of accretion disc are expected to be in a cold state and only slightly affect the process of viscous diffusion in  internal hot part of accretion flow.

\subsection{Typical time scales}
\label{sec:TimeScale}

\subsubsection{Time scales in {an} accretion disc}
\label{sec:TimeInTheDisc}

There are a few time scales which arise naturally from the physics of disc accretion onto {a} magnetized NS: 
(i) the dynamical time scale $t_{\rm K}$ given by Keplerian rotation, 
(ii) the viscous time scale $t_{\rm v}$, 
(iii) the time scale of {the} dynamo process, which provides the mechanism of viscosity in the accretion disc, and 
(iv) the time scale of instabilities $t_{\rm ins}$ at the inner disc radius, which are responsible for matter penetration into the NS magnetosphere.

The velocity of Keplerian rotation in accretion disc at radius $r$ is $v_\varphi=(GM/r)^{1/2}$ which corresponds to angular velocity 
$\Omega_{\rm K}=(GM/r^3)^{1/2}$. 
Thus, the dynamical timescale at the disc inner radius is given by
%\be
%\Omega_{\rm K}(R_{\rm m})=3.44\,\Lambda^{-3/2}B_{12}^{-6/7}L_{37}^{3/7}M_{1.4}^{2/7}R_6^{-15/7}\,{\rm rad\,s^{-1}} 
%\ee
%and the corresponding time scale is given by
\be\label{eq:TimeKepler}
t_{\rm K}(R_{\rm m})=\frac{2\pi}{\Omega_{\rm K}}\simeq 
1.9\,\Lambda^{3/2}B_{12}^{6/7}L_{37}^{-3/7}M_{1.4}^{-2/7}R_{\rm NS,6}^{15/7}\,{\rm s}.
\ee 

The viscous time scale $t_{\rm v}$ is determined by {the} radial velocity in {the} accretion disc $v_{\rm r}$, which is a result of viscous diffusion of accreting matter in the disc. The diffusion process is described by {the} viscous diffusion equation and depends on the kinematic viscosity $\nu$ (see Section \ref{sec:PropFluxTheory}).
In the simplified case of {an} $\alpha$-disc  \citep{1972AZh....49..921S,1973A&A....24..337S}, the radial velocity is given by
\be
v_{\rm r} \simeq \alpha v_\varphi \left(\frac{H}{R}\right)^2\ll v_\varphi,
\ee
where $0<\alpha<1$ is {the} dimensionless {viscosity} parameter, $H$ is {the} disc scale height {and} $c_{\rm s}$ is the isothermal sound speed.
The viscous time scale can be roughly estimated as
\be \label{eq:viscous_time}
t_{\rm v}(R)=\frac{t_{\rm K}(R)}{3\pi\alpha}\left(\frac{H(R)}{R}\right)^{-2}\gg t_{\rm K}(R).
\ee
In the C-zone of {an} accretion disc, the relative disc scale height can be estimated as \citep{2007ARep...51..549S}
\be\label{eq:H2R}
\frac{H(R)}{R}\approx 0.03\alpha^{-1/10}L_{37}^{3/20}m^{-21/40}R_{\rm NS, 6}^{3/20}R_8^{1/8},
\ee
where $R_8$ is the radial coordinate $R$ in accretion disc in units of $10^8\,{\rm cm}$.

Another important time scale of the problem is the typical timescale of initial fluctuations of the mass accretion rate in the disc $t_{\rm d}$. 
The fluctuations are likely caused by a magnetic dynamo {that} generates a poloidal field component in a random fashion and serves as the source of viscosity in the accretion flow, in the form of {correlated} fluctuations in magnetic stress
\citep{1991ApJ...376..214B,1995ApJ...440..742H,1995ApJ...446..741B,2018arXiv180904608L}. 
The dynamo timescale has been shown to be close to local Keplerian time-scale \citep{1992MNRAS.259..604T,1996ApJ...463..656S}: $t_{\rm d}\gtrsim t_{\rm K}$. 
Here, we assume 
\be \label{eq:TimeScaleMHD}
t_{\rm d}\approx k_{\rm d} t_{\rm K},
\ee 
where the exact value of $k_{\rm d}>1$ is not known \textit{a priori}, requiring detailed numerical simulations or/and detailed observational diagnostics.

These typical time scales determine the typical frequencies in the accretion disc. 
The Keplerian frequency:
$$
f_{\rm K}=t_{\rm K}^{-1}\simeq 0.53\,\Lambda^{-3/2}B_{12}^{-6/7}L_{37}^{3/7}M_{1.4}^{2/7}R_{\rm NS,6}^{-15/7}\,{\rm Hz}
$$
the viscous frequency:
$$
f_{\rm v}=t_{\rm v}^{-1}\simeq {0.94\,\alpha_{0.1} f_{\rm K}(R)}\left(\frac{H(R)}{R}\right)^{2}\ll f_{\rm K}(R),
$$
where $\alpha_{0.1}\equiv \alpha/0.1$, and the frequency of dynamo processes $f_{\rm d}=t_{\rm d}^{-1}$.
The introduced frequencies are related as
\be
f_{\rm K}\gtrsim f_{\rm d}\gg f_{\rm v}. 
\ee

\subsubsection{The time scale of emitting processes in the vicinity of {the} neutron star surface}

The geometry of {the} emitting region in the vicinity of {the} NS surface depends on the mass accretion rate and $B$-field strength \citep{2015MNRAS.447.1847M}.  
At low mass accretion rates (sub-critical, $\lesssim 10^{17}\,{\rm g\,s^{-1}}$), the accretion flow is braked by Coulomb collisions in the NS atmosphere \citep{1969SvA....13..175Z}. 
If the mass accretion rate is high enough (super-critical, $\gtrsim 10^{17}\,{\rm g\,s^{-1}}$), the accretion flow is stopped above NS surface by {a} radiation dominated shock \citep{1976MNRAS.175..395B,2015MNRAS.454.2539M}. Below the shock region, the accretion flow slowly settles to the stellar surface emitting energy in X-rays. Radiation {from the} accretion column can be beamed towards {the} NS surface \citep{1988SvAL...14..390L,2013ApJ...777..115P}. In this case, the photon energy flux is partly reprocessed/reflected by {the} NS atmosphere. The time scales of photon diffusion from a hot spot/accretion column and reprocessing of X-ray flux by the atmosphere are much shorter than the Keplerian time scale at the inner disc radius. 
Thus, these processes do not influence the PDS {in} the frequency range of our interest. 

At high mass accretion rates and relatively low $B$-field strength ($B<10^{12}\,{\rm G}$) at the NS surface, one would expect instabilities {in the} accretion column {to result in} in X-ray flares \citep{1976MNRAS.175..395B}. The time scale of the flares is expected to be in {the} range $10^{-4}<t_{\rm flares}<10^{-1}\,{\rm s}$.

\subsection{Accretion disc features in some classes of X-ray pulsars}
\label{sec:AD_in_classes}

\subsubsection{Accretion discs in transients}
\label{sec:ColdDisc}

Outbursts of X-ray transients are likely driven by the development of a disc instability 
triggered by gradual accumulation of matter in the disc or episodic material capture from a companion star (like it happens in Be/X-ray transients, see e.g. \citealt{2011Ap&SS.332....1R}) 
and caused by the strong dependence of viscosity on temperature {for temperatures in the range} $T\sim 6500\,{\rm K}$ (see e.g. \citealt{2001NewAR..45..449L}). 
The development of {this} instability results in cooling and heating waves propagating in the accretion disc. The physical conditions (particularly, the viscosity) in the disc are significantly different at opposite sides of the cooling/heating waves.
It {is} likely that only {the} "hot" part of {the} accretion disc effectively contributes to the propagating fluctuations in mass accretion rate (the initial fluctuations are smaller in weakly ionized parts of {the} disc and suppression of variability at high frequencies is stronger in a cold part of {the} accretion disc because of the much longer viscous time scale). 
We expect that the evolution of the outburst and dynamics of cooling/heating waves in accretion disc can affect the PDS of aperiodic variability and even cause QPOs at low frequencies. 
However, these effects are beyond the scope of the paper and will be considered separately.

\subsubsection{Accretion discs in ULX pulsars}
\label{sec:ULXs}

It has been recently discovered that the luminosity of XRPs can exceed hundreds of Eddington luminosities \citep{2014Natur.514..202B,2017Sci...355..817I} and at least a fraction of ULXs host accreting NSs.
The exact mechanism of such an extreme mass accretion rate is still under debate \citep{1976MNRAS.175..395B,1992AcA....42..145P,2015MNRAS.454.2539M,2017MNRAS.467.1202M,2017MNRAS.468L..59K}, but it is clear that the conditions of accretion discs in these systems are different from those in normal XRPs. {In} particular, one would expect a geometrically thick inner part of the accretion disc, where the pressure is dominated by radiation (see A-zone in Fig.\,\ref{pic:DiscZones}). The thick inner part of {the} disc can affect the timing properties of aperiodic variability due to different radial dependence of the kinematic viscosity $\nu$ and, therefore, different transfer properties of propagating fluctuations. {A} radiation pressure dominated region and possible mass losses from {the} accretion disc \citep{1999AstL...25..508L,2007MNRAS.377.1187P} can also result in a dependence of the inner disc radius on the mass accretion rate different from the one given by equation \ref{eq:Rm} (see Tab.\,1 in \citealt{1999ApJ...521..332P}, see also \citealt{2017MNRAS.470.2799C,2019arXiv190204609C,2019MNRAS.484..687M}).

Another feature of magnetized NSs at extremely high mass accretion rates is {an} optically thick envelope formed by {the} accretion flow moving from magnetospheric radius to the central object \citep{2017MNRAS.467.1202M}. The envelope hides the NS from a distant observer and affects the observational manifestation of ULX pulsars including their fast aperiodic variability. Particularly, the envelope suppresses high-frequency variability of {the} X-ray energy flux \citep{2019MNRAS.484..687M}.

\section{Propagating fluctuations of the mass accretion rate}
\label{sec:PropFluxTheory}

In this Section, we develop a model for propagating accretion rate fluctuations in an XRP disc with inner radius at the magnetospheric radius, $R_{\rm in}=R_m$, and outer radius at the tidal radius, $R_{\rm out}=R_{\rm tid}$. 
We consider small amplitude local fluctuations of mass accretion rate/surface density arising at each radius in the accretion disc and propagating due to the process of viscous diffusion \citep{2018MNRAS.474.2259M}. The fluctuations propagate both inwards and outwards (Mushtukov et al. 2018) contributing to the total variability of the mass accretion rate at each radial coordinate. The propagation of fluctuations is governed by the viscous diffusion equation (Section \ref{sec:ViscousDiffEquation}). In Section 3.2, we present our solution to the diffusion equation that, for the first time, accounts for a finite inner and outer radius which we will use to model the PDS of XRPs. In Section 3.3, we explore the properties of our new Green function and in Section 3.4, we define our model for the power spectrum of initial fluctuations assumed to be injected into the disc.

\subsection{The viscous diffusion equation}
\label{sec:ViscousDiffEquation}

Propagation of the mass accretion rate fluctuations can be accurately described by the solutions of the equation of viscous diffusion \citep{1974MNRAS.168..603L}:
\be\label{eq:DifEqGen}
\frac{\partial \Sigma(R,t)}{\partial t}=\frac{1}{R}\frac{\partial}{\partial R}\left[R^{1/2}\frac{\partial}{\partial R}\left(3\nu \Sigma R^{1/2}\right)\right],
\ee
where $\Sigma$ is the local surface mass density, $R$ is the radial coordinate, $\nu$ is the kinematic viscosity, and $t$ is time.
In the particular case {whereby} kinematic viscosity {is not} dependent on the local surface density, the equation of viscous diffusion is linear. 
Then the solutions of the equation can {then} be found using the Green functions:
\be
\Sigma(R,t)=\int\limits_{R_{\rm in}}^{R_{\rm out}}G(R,R',t-t_0)\Sigma(R',t_0)\d R' ,
\ee
where $G(R,R',t)$ is a Green function describing evolution of the surface density, $\Sigma(R,t_0)$ is the surface density distribution over the radial coordinate $R$ at $t=t_0$, and $R_{\rm in}$ and $R_{\rm out}$ are inner and outer disc radius respectively. 

The particular Green functions are determined by viscosity dependence on the radial coordinate $\nu(R)$ and boundary conditions at the inner $R_{\rm in}$ and outer $R_{\rm out}$ radii of the disc.
The solution to the equation of viscous diffusion simplifies if we assume {a} power-law dependence of kinematic viscosity on radial coordinate:
\be\label{eq:nu(R)}
\nu(R)=\nu_0 \left( \frac{R}{R_0} \right)^n.
\ee
In zone-C of the \cite{1973A&A....24..337S} disc model, which is relevant to the XRP discs we consider here, we have:
\be
\nu=\alpha\Omega_{\rm K}R^2 \left(\frac{H}{R}\right)^2\propto R^{3/4}, 
\ee
i.e. $n=3/4$ (assuming constant $\alpha$). The assumption (\ref{eq:nu(R)}) is applicable for a wide range of mass accretion rates and radii if we consider relatively small oscillations of accretion rate around an average value. Under assumption (\ref{eq:nu(R)}), Green functions have been derived analytically for a few particular cases:
(1) for the case of $R_{\rm in}=0$, $R_{\rm out}=\infty$ the Green functions was derived by \cite{1974MNRAS.168..603L};
(2) for the case of $R_{\rm in}>0$, $R_{\rm out}=\infty$ the Green function was derived by \cite{2011MNRAS.410.1007T};
(3) for the case of $R_{\rm in}=0$, $R_{\rm out}<\infty$  and zero mass accretion rate at $R_{\rm out}$ the Green function was derived by \cite{2015ApJ...804...87L}.
The zero torque/zero mass accretion rate condition at $R_{\rm in}$ was adopted in each of the mentioned works.
Green functions accounting for the effects of general relativity in the Kerr geometry have been derived by \cite{2017MNRAS.471.4832B}.
 In this paper, we present a Green function solution to equation (\ref{eq:DifEqGen}) that accounts for $R_{\rm in}>0$ and $R_{\rm out}<\infty$, and can therefore be used to describe XRP discs.

Variability of the surface density is accompanied by variability of local mass accretion rate:
\beq\label{eq:mdot00}
\dot{M}(R,t)=6\pi R^{1/2}\frac{\partial}{\partial R}\left(\nu\Sigma(R,t)R^{1/2}\right).
\eeq
Thus, the mass accretion rate variability can be represented as 
\beq\label{eq:mdot01_}
\dot{M}(R,t)=\int\limits_{R_{\rm in}}^{R_{\rm out}}G_{\dot{M}}(R,R',t)\otimes_t \frac{\partial \Sigma^*(R',t)}{\partial t}\d R' ,
\eeq
where $\otimes_x$ denotes the convolution in $x$-variable,
$\Sigma^*$ is {the initial fluctuation of the surface density due to local MHD processes},
and 
$G_{\dot{M}}(R,R',t)$ is the Green function for the mass accretion rate:
\be 
\label{eq:mdot02}
G_{\dot{M}}(R,R',t)=6\pi R^{1/2}\frac{\partial}{\partial R}\left(\nu G(R,R',t)R^{1/2}\right).
\ee
Equation (\ref{eq:mdot02}) gives the Green function in the time domain, but in the analysis of timing properties of mass accretion rate fluctuations it is convenient to use the Green function in the frequency domain.

The Green function in the frequency domain is given by {the} Fourier transform:
\be\label{eq:FT}
\overline{G}_{\dot{M}}(R,R',f)=\int\limits_{-\infty}^{\infty}\d x\,G_{\dot{M}}(R,R',x)\,e^{-2\pi i f x} 
\ee
and has the physical meaning of {a} transfer functions {that} describes how the variability at radial coordinate $R'$ affects variability at the the radial coordinate $R$ (see e.g. \citealt{2018MNRAS.474.2259M}). 
The Green function in the frequency domain belongs to the complex plane and contains information both about the amplitude and the time delay of the transferred fluctuations.
The PDS of mass accretion rate fluctuations at radius $R$ can be calculated as
\beq
\label{eq:S_mdot}
S_{\dot{M}}(R,f)\simeq \int\limits_{R_{\rm in}}^{R_{\rm out}}\frac{\d R'}{(R')^2}\Delta R(R')
|\overline{G}_{\dot{M}}(R,R',f)|^2 S_{a}(R',f),
\eeq
where $S_{a}(R',f)$ is the PDS of initial fluctuations, $\Delta R(R)$ is a typical radial scale {on which} the initial fluctuations can be considered coherent, and the integral is taken over the whole range of radial coordinates in {the} accretion disc. 
It is likely that $\Delta R(R)\sim H(R)$ \citep{2016ApJ...826...40H} and we use the estimation given by (\ref{eq:H2R}) in our numerical calculations.
Thus, the PDS of the mass accretion rate at any radius (including the inner disc radius) is entirely determined by the PDS of initial fluctuations (see Section \ref{sec:IniFlux}) and the transfer functions given by suitable Green functions in the frequency domain (see Sections \ref{sec:GfTanaka} and \ref{sec:GfLipunova}).
Note that the expressions introduced in this section do not account for non-linear effects arising because of interaction of propagating fluctuations with each other.
This approximation is good in the case of a small fractional rms \citep{2018MNRAS.474.2259M}.

Propagation of fluctuations of the mass accretion rate under {the} influence of viscous diffusion suppresses high-frequency ($f>f_{\rm v}$) variability \citep{2001MNRAS.327..799K,2018MNRAS.474.2259M}. The initial variability at a given radius in {the} accretion disc affects mass accretion rate variability both in the inner and outer part of {the} disc \citep{2018MNRAS.474.2259M}.

\subsection{New Green function for {an} accretion disc with finite inner and outer radii}
\label{sec:GfLipunova}

The \cite{2011MNRAS.410.1007T} Green function solution to equation (\ref{eq:DifEqGen}) is reproduced in Appendix \ref{sec:GfTanaka}. 
This solution accounts for a finite disc inner radius but assumes $R_{\rm out}=\infty$ and is therefore not appropriate for modelling XRP accretion discs, for which the disc’s inner radius is not much smaller than its outer radius. 
We therefore derive a new Green function solution to equation (\ref{eq:DifEqGen}) with finite $R_{\rm in}$ and $R_{\rm out}$ (Appendix \ref{sec:G_F}), which we will use for our subsequent analysis of XRP PDSs. 
We assume zero mass accretion rate at the outer radius, $\dot{M}(R_{\rm out})=0$, implying that the disc loses mass from the inner radius only. 
This is appropriate for $R_{\rm out}$ set by the tidal radius or by the boundary between the hot and cold parts of the disc \citep{2017MNRAS.468.4735L}. 
Note, that mass inflow from the companion star is possible at any radial coordinate within the outer disc radius. 
The exact geometry of the mass inflow is determined by {the} geometry of {the} binary system and {the} mechanism of mass transfer (wind accretion, accretion from the Lagrangian point or matter capture during the periastron passage). 
We will first assume zero torque at the inner radius (Section \ref{sec:ZeroTorque}), before considering general torque at the inner radius (Section \ref{sec:NonZeroTorque}). 
Remarkably, we find that our Green function derived assuming zero torque at $R_{\rm m}$ can be generally applied to XRP discs.

\subsubsection{The case of a zero-torque inner boundary}
\label{sec:ZeroTorque}

In this limit, our Green function is (see Appendix\,\ref{sec:G_F} and Fig.\,\ref{pic:GreenLipEx00}a):
\beq\label{eq:GF_LipSigma}
&G(R,R',t)= (2-n) R^{-n-1/4} R'^{5/4}R_{\rm out}^{n-2} \\
&\times\sum\limits_{i}
\exp\left[-2\left(1-\frac{n}{2}\right)^2 k_i^2\frac{t}{t_{\rm v}}\right]\frac{V_l(k_i x',k_i x_{\rm in})V_l(k_i x,k_i x_{\rm in})}{V_l^2(k_i x_{\rm out},k_i x_{\rm in})}, \nonumber
\eeq
where $x=(R/R_{\rm out})^{1-n/2}$, 
\be\label{eq:V} 
V_l(u,\nu)=J_l(u)J_{-l}(\nu)-J_{-l}(u)J_{l}(\nu),
\ee
$J_l(x)$ are the Bessel functions of the first order,
and $k_i$ are roots of {the} transcendental equation 
\beq 
k_i [J_{l-1}(k_i x_{\rm out})J_{-l}(k_i x_{\rm in})-J_{-l-1}(k_i x_{\rm out})J_l(k_i x_{\rm in})]  \nonumber \\
-\frac{2l}{x_{\rm out}}J_{-l}(k_i x_{\rm out})J_l(k_i x_{\rm in})=0. \nonumber
\eeq

\begin{figure}
\centering 
\includegraphics[width=9.cm]{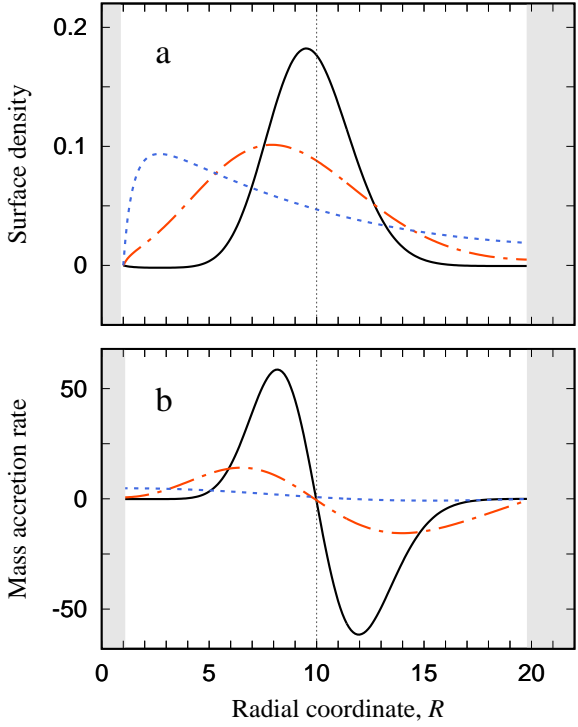}
\caption{
{Examples of Green functions describing (a) viscous evolution of a perturbation in surface density originating from $R'=10$ and 
(b) the corresponding perturbation in mass accretion rate, both for a disc truncated from inside at $R_{\rm in}=1$ and outside at $R_{\rm out}=20$.
Different curves correspond to different moments {in} time: $t=0.04\,t_{\rm in}$ (solid black), $t=0.16\,t_{\rm in}$ (dashed-dotted red), $t=0.64\,t_{\rm in}$ (dashed blue), where $t_{\rm in}$ is the viscous time at $R=R_{\rm in}$ (see equation \ref{eq:viscous_time}).}
Parameters: $n=0.75$.
}
\label{pic:GreenLipEx00}
\end{figure}

Using (\ref{eq:mdot02}) and (\ref{eq:FT}) we get the Green function for the mass accretion rate in the frequency domain (see Fig.\ref{pic:GreenLipEx01} and \ref{pic:GreenLipEx02}):
\beq\label{eq:GF_Lip_Mdot}
 \overline{G}_{\dot{M}}(R,R',f)=6\pi R^{1/2} \left(\frac{R_{\rm out}}{R_{\rm in}}\right)^{2-n} \nonumber \\
 \times\frac{\partial}{\partial R}\left\{
 R^{n+0.75}\sum\limits_{i}\frac{V_l(k_i x',k_i x_{\rm in})V_l(k_i x,k_i x_{\rm in})}{V_l^2(k_i x_{\rm out},k_i x_{\rm in})} \right. \nonumber\\
 \left.
\times\frac{1}
 {4\pi i f/f_{\rm in}+k_i^2(n-2)^2}
 \right\}.
\eeq
The expression (\ref{eq:GF_Lip_Mdot}) can be reduced to 
\beq\label{eq:GF_Lip_fr}
& \overline{G}_{\dot{M}}(R,R',f)=\frac{6\pi}{4} (2-n) R^{-1/4} \left(\frac{R_{\rm out}}{R_{\rm in}}\right)^{2-n} \nonumber \\
& \times
 \sum\limits_{i}\frac{V_l(k_i x',k_i x_{\rm in})W_l(k_i x,k_i x_{\rm in})}{V_l^2(k_i x_{\rm out},k_i x_{\rm in})}
\frac{1}
 {4\pi i f/f_{\rm in}+k_i^2(n-2)^2},
\eeq
where
\beq
W_l(k_i x,k_i x_{\rm in})&=&
V_l(k_i x,k_i x_i) \nonumber \\
&&+J_{-l}(k_i x_i)[J_{l-1}(k_i x)-J_{l+1}(k_i x)] \nonumber \\
&&-J_{l}(k_i x_i)[J_{-l-1}(k_i x)-J_{-l+1}(k_i x)], \nonumber
\eeq
and $f_{\rm in}=f_{\rm v}(R_{\rm in})$ is the viscous frequency at the inner disc radius.
Equation (\ref{eq:GF_Lip_fr}) gives {an} analytic expression for the Green function in the frequency domain.
{In the limiting case of $R_{\rm out}\rightarrow \infty$, the expression (\ref{eq:GF_Lip_fr}) turns to (\ref{eq:GfTanaka_f}), as expected.

\begin{figure}
\centering 
\includegraphics[width=9.3cm]{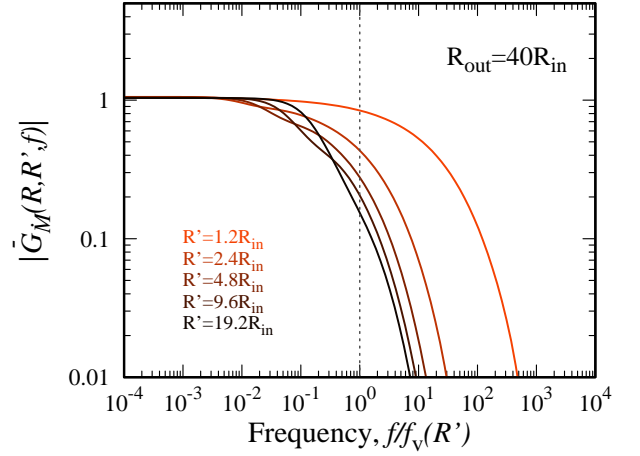}
\caption{
The absolute value of the transfer functions $|G_{\dot{M}}(f,R,R')|$ calculated for radial coordinate $R=R_{\rm in}=1$ and $R'=1.2,\,2.4,\,4.8,\,9.6,\,19.2\,R_{\rm in}$. 
Note that the frequency is measured here in units of the local viscous frequency corresponding the radial coordinate of initial perturbations $R'$.
Only the fluctuations aroused close to the inner disc radius propagate inwards without a significant suppression of variability at the frequencies above the local viscous frequency $f_{\rm v}(R')$ (vertical dashed line).
Parameters: $R_{\rm in}=1$, $R_{\rm out}=40$, $n=0.75$.
}
\label{pic:GreenLipEx01_}
\end{figure}

\begin{figure}
\centering 
\includegraphics[width=9.3cm]{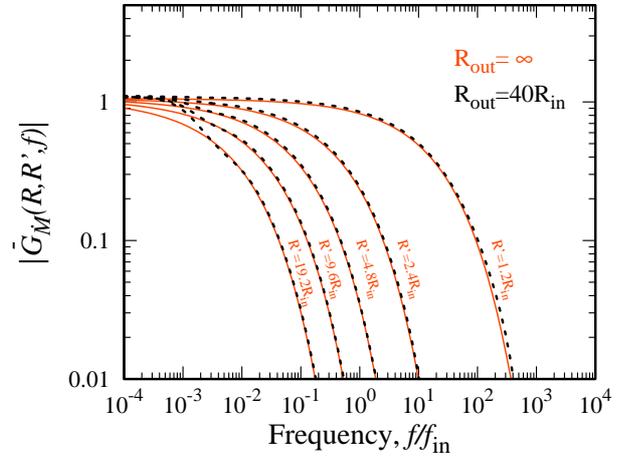}
\caption{The absolute value of the transfer functions $|G_{\dot{M}}(f,R,R')|$ calculated for radial coordinate $R=R_{\rm in}=1$ and $R'=1.2,\,2.4,\,4.8,\,9.6,\,19.2\,R_{\rm in}$. 
The red solid lines correspond to {an} infinite accretion disc $R_{\rm out}=\infty$ (see Section \ref{sec:GfTanaka}), while the black dashed lines correspond to {a} disc {with} outer radius $R_{\rm out}=40$ (see Section \ref{sec:GfLipunova}). There is a difference at low frequencies when $R'$ becomes close to the outer disc radius and the diffusion process "feels" the outer boundary of the accretion disc.
Parameters: $R_{\rm in}=1$, $n=0.75$.
}
\label{pic:GreenLipEx01}
\end{figure}

\begin{figure}
\centering 
\includegraphics[width=9.cm]{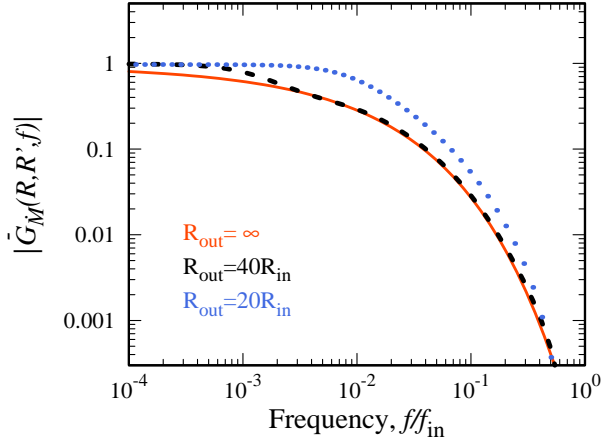}
\caption{The Green functions for the mass accretion rate in the frequency domain calculated for $R=R_{\rm in}=1$ and $R'=19.2\,R_{\rm in}$ in accretion discs of different outer radii: $R_{\rm out}=\infty$ (red solid line), $R_{\rm out}=40 R_{\rm in}$ (black dashed line) and $R_{\rm out}=20 R_{\rm in}$ (blue dotted line).
}
\label{pic:GreenLipEx02}
\end{figure}

\begin{figure}
\centering 
\includegraphics[width=8.5cm]{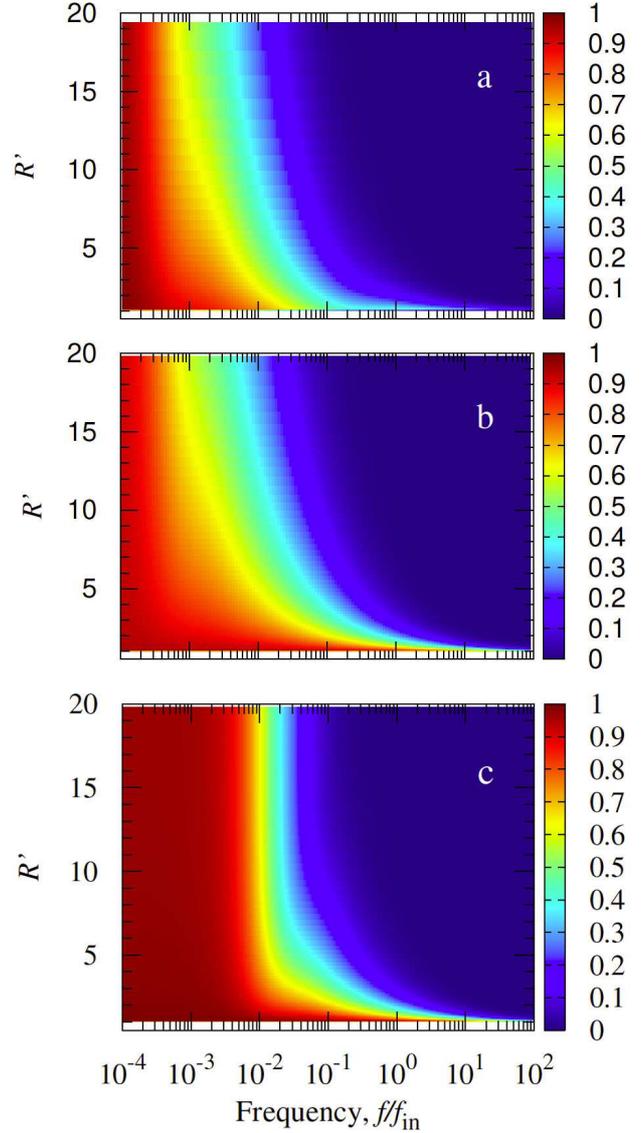}
\caption{The absolute value of the transfer function $|G_{\dot{M}}(f,R=1,R')|$ for the cases of 
(a) {an} infinite accretion disc {with} $R_{\rm in}=0$ (described by Lynden-Bell Green functions);
(b) {an} infinite accretion disc truncated at $R_{\rm in}=1$ (described by Tanaka's Green functions);
(c) {an} accretion disc {with} outer radius $R_{\rm out}=20$ truncated at $R_{\rm in}=1$. 
One can see that accretion discs truncated from {the} inside ($R_{\rm in}\ne 0$) keep {more} high frequency variability originating from the radii adjacent to the given radius
(compare (a) with (b,c));
accretion discs truncated from the outside keep some extra variability from the outer disc parts (compare (c) with (a,b)).  
Parameters: $n=0.75$, $R_{\rm in}=1$.
}
\label{pic:GreenLipEx03}
\end{figure}

\subsubsection{The case of a non-zero-torque inner boundary}
\label{sec:NonZeroTorque}

Whilst a zero-torque inner boundary condition may be typical for accretion discs around black holes, in the case of accretion onto magnetised NSs, the torque at the inner disc radius {may have} a finite value, which is determined by the the spin-up rate of {the} rotating magnetized NS.
Stable accretion onto {a} magnetised NS is possible if the the mass accretion rate is high enough {for} the accretion flow {to} penetrate through the centrifugal barrier set up by the rotating magnetosphere \citep{1975A&A....39..185I}.
Non-zero torque at the inner disc radius affects the torque distribution all over the disc.

Viscous torques in accretion disc are determined by {the} local surface density and viscosity.
If the viscosity depends on radial coordinate only,  the linear dependence between the local viscous torque $F$ and surface density $\Sigma$ holds:
\be
F(R,t)=3\pi h \nu_0\Sigma(R,t)\left(\frac{R}{R_0}\right)^n ,
\ee
where $h=(GMR)^{1/2}$ is the specific angular momentum.
The local mass accretion rate is determined by the derivative 
\be\label{eq:F2dotM}
\dot{M}(R,t)=\frac{\partial F(R,t)}{\partial h}. 
\ee
In the case of  stationary accretion in a disc with a non-zero torque at the inner radius, the the viscous torques in {the} disc $F$ are given by
\be\label{eq:F0}
F(R)=F_{\rm in}+\int\limits_{h_{\rm in}}^{h}\d h\,\dot{M}(R), 
\ee
where $F_{\rm in}$ is the torque at the inner disc radius, 
i.e. non-zero torque at the inner disc radius results in {an} increase of viscous torques all over the disc by a constant value. 
The time dependent torque can be represented as {the} sum of {the} time independent torque corresponding to stable accretion (\ref{eq:F0}) and {a} fluctuating viscous torque $F_2$ on top of it:
\be \label{eq:F01}
F(R,t)=F_{\rm in}+\int\limits_{h_{\rm in}}^{h}\d h\,\dot{M}_0(R)+F_2(R,t),
\ee
where $\dot{M}_0$ is the average mass accretion rate. 
The first two terms {on} the right hand side of (\ref{eq:F01}) represent the time independent solution of the equation of viscous diffusion.
Because the equation of viscous diffusion is considered to be linear, the fluctuating part of the solution is a solution of the viscous diffusion equation by itself. 
Note that $F_2(R,t)|_{R=R_{\rm in}}=0$ and, therefore, it satisfies the equation of viscous diffusion with zero torque boundary condition at $R_{\rm in}$. 
According to (\ref{eq:F2dotM}) the time dependent mass accretion rate can be obtained from (\ref{eq:F01}):
$$
\dot{M}(R,t)=\dot{M}_0 +  \frac{\partial F_2(R,t)}{\partial h},
$$
where the second term represents local fluctuations of the mass accretion rate on top of the average $\dot{M}_0$.
Thus, the fluctuations of the mass accretion rate
$\dot{m}(R,t)={\partial F_2(R,t)}/{\partial h}$
are described by {a} Green function derived for zero torque at the inner disc radius (see Section \ref{sec:ZeroTorque}).

\subsection{Properties of the transfer functions of propagating fluctuations}
\label{sec:Properties}

Fig.\,\ref{pic:GreenLipEx00} shows an example of our Green function for the surface density (top; Equation \ref{eq:GF_LipSigma}) and mass accretion rate (bottom; Equation \ref{eq:GF_Lip_Mdot}). 
We see that an initial perturbation in the surface density at $R'=10$ spreads out over time, with more material propagating inwards than outwards. 
This creates a perturbation in the accretion rate that is initially positive inside $R'=10$ and negative outside of $R'=10$ before the accretion rate slowly settles back to its equilibrium value (zero in the Figure). 
In the frequency domain, the Green functions of the mass accretion rate $\overline{G}_{\dot{M}}(R,R',f)$ play the role of transfer functions {that} describe how variability at Fourier frequency $f$ {and} radial coordinate $R'$ in the disc affects variability at the coordinate $R$, see equation (\ref{eq:S_mdot}). 
In particular, the absolute value of the transfer function demonstrates {the} survival level of initial variability: 
$|\overline{G}_{\dot{M}}(R,R',f)|=1$ corresponds to the case of fluctuations propagating without suppression, 
while $|\overline{G}_{\dot{M}}(R,R',f)|\ll1$ corresponds to significant suppression of initial variability. 
In this paper, we are interested in mass accretion rate variability at the inner disc radius and focus on the inward propagation of fluctuations (i.e. $R<R'$).

The process of viscous diffusion effectively suppresses variability at frequencies $f\gtrsim f_{\rm v}(R')$, where $f_{\rm v}(R')$ is a local viscous frequency corresponding to the radial coordinate of initial fluctuations, unless the radial coordinate of initial fluctuations is very close to the inner disc radius (see Fig.\,\ref{pic:GreenLipEx01_}).
Fluctuations arising close to the inner disc radius can propagate inwards without significant suppression.
Fig.\,\ref{pic:GreenLipEx01} shows that the effect of finite $R_{\rm out}$ becomes important for the propagation of fluctuations that originated at large radii, and Fig.\,\ref{pic:GreenLipEx02} shows that the effect is more pronounced for smaller $R_{\rm out}$ (as one would expect). 
Fig.\,\ref{pic:GreenLipEx03} compares the fluctuations for three different Green function solutions: (a) $R_{\rm in}=0$ and $R_{\rm out}=\infty$, (b) $R_{\rm in}>0$ and $R_{\rm out}=\infty$, and (c) $R_{\rm in}>0$ and $R_{\rm out}<\infty$. We see that the boundary conditions affect the effectiveness of propagation of mass accretion rate fluctuations {in the following ways} (see Fig.\ref{pic:GreenLipEx03}):

(i) Variability originating near the non-zero inner radius of the disc is weakly supressed and the inner disc regions contribute more high-frequency variability comparing to the case with $R_{\rm in}=0$
(compare Fig.\ref{pic:GreenLipEx03}a and Fig.\ref{pic:GreenLipEx03}b for $f/f_{\rm v} > 10^{-1}$), and thus the inner disc regions contribute more high-frequency variability at the inner radius {in the former case}. This happens because there is no flow of  angular momentum to $R_{\rm in}$ from $R<R_{\rm in}$ and the time scale of radial transfer becomes smaller.

(ii) The accretion discs {with} outer radius $R_{\rm out}<\infty$ with $\dot{M}(R_{\rm out})=0$ preserve some extra variability originating from the outer parts of accretion disc (see Fig.\ref{pic:GreenLipEx02}, compare Fig.\ref{pic:GreenLipEx03}b and Fig.\ref{pic:GreenLipEx03}c for $f/f_{\rm v}<10^{-2}$).

\subsection{Power spectra of initial perturbations}
\label{sec:IniFlux}

An essential ingredient of the model predicting timing properties of broadband aperiodic variability is the properties of initial fluctuations produced in the accretion disc. 
Starting from {given} properties of the initial fluctuations, we can describe their transfer properties using suitable Green functions of the viscous diffusion equation (see equation \ref{eq:S_mdot}) {to obtain} predictions on the aperiodic variability in X-rays. 
However, the timing properties of the initial perturbations of the surface density and mass accretion rate are not known precisely. 
As discussed in Section \ref{sec:TimeInTheDisc}, they may be caused by a magnetic dynamo process \citep{1991ApJ...376..214B,1995ApJ...440..742H,1995ApJ...446..741B}, which has typical timescale 
$t_{\rm d}\approx k_{\rm d}f_{\rm K}^{-1}$, where $k_{\rm d}\sim few$ \citep{1992MNRAS.259..604T}. 
Here, we specify the PDS of initial fluctuations with a Lorentzian function
\footnote{ 
The full information about the timing properties of initial perturbations is given by {the} cross-spectrum of initial variability at various radial coordinates in {the} accretion disc \citep{2018MNRAS.474.2259M}.
However, one can use {the} PDS of initial fluctuations $S_a(f)$ instead of {the} cross-spectrum in order to get {an} approximate solution (see equation \ref{eq:S_mdot}).
The approximation is good if different rings in {the} accretion disc produce uncorrelated initial perturbations of the mass accretion rate.} 
\be\label{eq:LorentzProf_gen}
S_{a}(R,f) = 
\frac{{ F_{\rm var} }/{ (\ln 10~R) }}  {\arctan(f_{br}/f_0)}
\frac{f_{\rm br}(R)}{(f_{\rm br}(R))^2+(f-f_0)^2},
\ee
where $F_{\rm var}$ is the fractional variability amplitude per radial decade generated by turbulence in the disc (following e.g. \citealt{2006MNRAS.367..801A,2011MNRAS.415.2323I,2012MNRAS.419.2369I,2013MNRAS.434.1476I}). 
This means that the integrated power of the perturbations
$P(R)=\int_0^\infty S_a(f) df = { F_{\rm var} }/{ (\ln 10 ~ R) }\propto 1/R$ 
for a constant $F_{\rm var}$, which is expected from MRI turbulence in a constant $h/r$ disc (e.g. \citealt{2001MNRAS.321..759C,2006MNRAS.367..801A}). 
Since we are ignoring the influence of non-linear effects on the shape of the power spectrum, the $F_{\rm var}$ model parameter acts as a simple normalisation of the predicted XRP power spectrum.

The frequencies $f_0$ and $f_{\rm br}$ are model parameters, and we assume $f_{\rm br}(r) = f_{\rm d}(r) \propto f_{\rm K}(r)/k_{\rm d}$, where the dynamo coefficient $k_{\rm d}$ is a model parameter. 
The choice of Lorentzians is motivated by the fact that {an} exponentially decaying periodic signal 
$$g(t)=\sin(2\pi f_0 t)e^{-t/T}$$
is described by 
$$\overline{g}(f)\propto [(f-f_0)^2+(1/T)^2]^{-1/2}$$ 
in the frequency domain, which gives the Lorentzian {when squared}.

\section{Modeling the Power Density Spectra}
\label{sec:NumResults}

{
Fluctuations of X-ray energy flux in XRPs are determined by fluctuations of the mass accretion rate at the inner edge of the accretion disc. 
Therefore, we model {the} PDS of fluctuations of the mass accretion rate at the inner disc radius and compare it with the PDS of observed fluctuations of X-ray energy flux. 
We use dimensionless radial coordinates and dimensionless frequencies in our calculations, i.e. the radial coordinate and the frequency are measured in units that set the physical scales in the system. 
In particular, the frequency is measured in units of {the} viscous frequency {at the disc inner radius}, which is related to viscous and geometrical properties of {the} accretion disc (see Section \ref{sec:TimeInTheDisc}). 
}

\subsection{The major physical parameters and examples of PDS}

The PDS of the mass accretion rate at the inner disc radius is determined by a set of parameters:
(i) the inner and outer radii of {the} accretion disc and boundary conditions there (Fig.\,\ref{pic:GreenLipEx05});
(ii) {the} PDS of initial fluctuations (Fig.\,\ref{pic:GreenLipEx04});
(iii) {the} dependence of the kinematic viscosity on the radial coordinate in the disc (Fig.\,\ref{pic:GreenLipEx06});
(iv) {the} dependence of total power of initial fluctuation on the radial coordinate in the disc (Fig.\,\ref{pic:GreenLipEx07}).
{We see that} the broadband variability covers a few orders of magnitude in the frequency domain. 
The typical theoretical PDS has two breaks. 
The break at lower frequencies corresponds to the viscous frequency $f_{\rm v}$ at the outer disc radius, while the break at high frequencies corresponds to the typical frequencies of initial fluctuations at the inner disc radius (see Fig.\,\ref{pic:GreenLipEx05}). 
Both breaks are smooth and can cover an order of magnitude in frequency, which poses difficulties for interpretation of observational results. 
The exact shape of {the} break at high frequency is strongly affected by the shape of {the} PDS of initial fluctuations (see Fig.\,\ref{pic:GreenLipEx04}).

The decrease of the inner disc radius (which can be caused by increasing mass accretion rate in XRPs, see equation \ref{eq:Rm}) results in a shift of the high-frequency break to even higher frequencies and a decrease of {the} PDS at frequencies below the high-frequency break (compare black solid and blue dashed-dotted lines in Fig.\,\ref{pic:GreenLipEx05}). 
{The shift of the break frequency is caused by the inclusion of a new inner part of {the} disc {that} is able to produce fast variability at time scales not achievable in {an} accretion disc {with} larger inner radius. 
The same inner part of the accretion disc additionally suppresses variability from the outer parts. It leads to {a} decrease of the PDS at low frequencies (compare black and blue lines of Fig.\,\ref{pic:GreenLipEx05} in the frequency range $10^{-2}\lesssim f/f_{\rm v}(R_{\rm m}) \lesssim 10^{1}$). 
The PDS above {the} high-frequency break deviates from {a} power-law.}

\begin{figure}
\centering 
\includegraphics[width=9.cm]{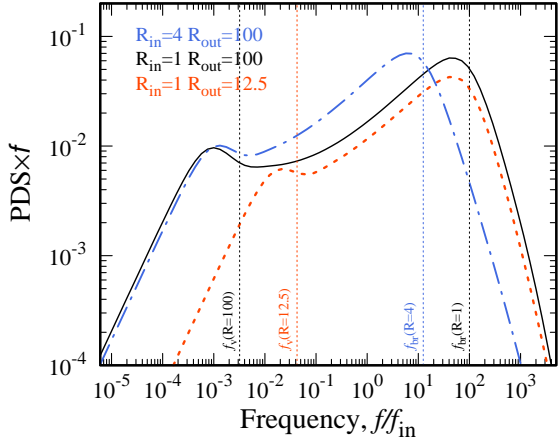}
\caption{ 
The PDS of the mass accretion rate variability at $R_{\rm in}$ calculated for accretion discs of different inner and outer radii:
(i) $R_{\rm in}=1$, $R_{\rm out}=100$ (black solid line),
(ii) $R_{\rm in}=1$, $R_{\rm out}=12.5$ (red dashed line),
(iii) $R_{\rm in}=4$, $R_{\rm out}=100$ (blue dashed-dotted line).
The low-frequency break in {the} PDS corresponds to the viscous frequency at $R_{\rm out}$, while the high-frequency break corresponds to the break frequency of initial fluctuations at the inner part of the disc. 
Both breaks are quite broad.  
Parameters: $n=0.75$, $f_{\rm br}=100R^{-3/2}$.
}
\label{pic:GreenLipEx05}
\end{figure}

\begin{figure}
\centering 
\includegraphics[width=9.cm]{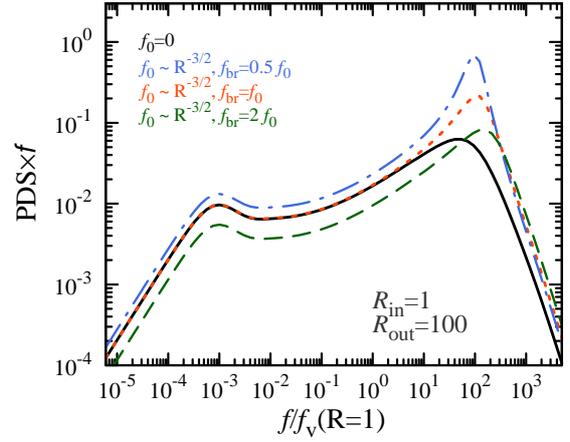}
\caption{ 
The PDS of mass accretion rate variability at $R_{\rm in}$. Different curves are given for different PDSs of the initial fluctuations: 
zero-centered Lorentzians breaking at frequency $f_{\rm br}=100R^{-3/2}$ (black solid line) and
Lorentzians of various width $f_{\rm br}$ centered at $f_0$ (see equation \ref{eq:LorentzProf_gen}).
Parameters: $n=0.75$.
}
\label{pic:GreenLipEx04}
\end{figure}

\begin{figure}
\centering 
\includegraphics[width=9.cm]{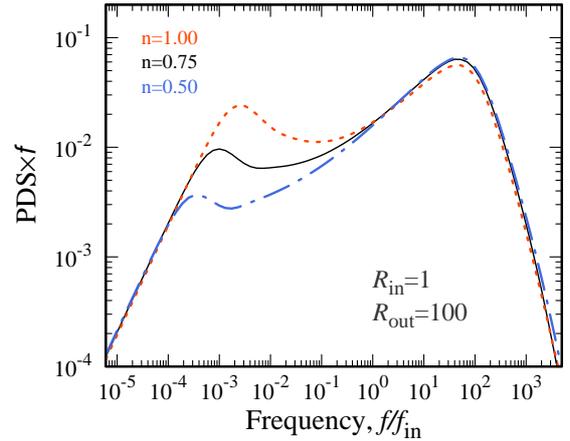}
\caption{ 
The PDS of mass accretion rate variability at $R_{\rm in}$. Different curves correspond to different powers $n$ in the dependence of kinematic viscosity on the radial coordinate in accretion disc: 
$n=0.5$ (blue dashed-dotted), 
$n=0.75$ (black solid), 
$n=1$ (red dashed).
Parameters: $f_{\rm br}=100R^{-3/2}${, $f_0=0$}.
}
\label{pic:GreenLipEx06}
\end{figure}

\begin{figure}
\centering 
\includegraphics[width=9.cm]{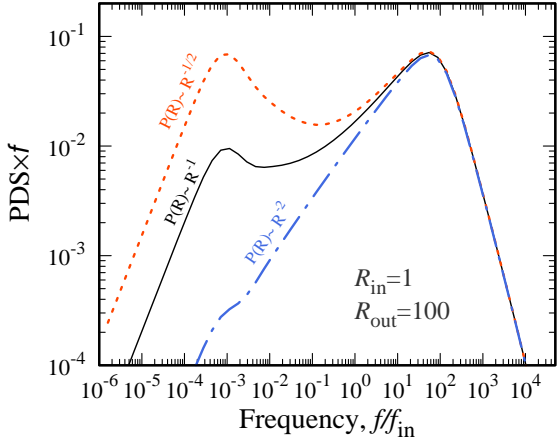}
\caption{ 
The PDS of mass accretion rate variability at $R_{\rm in}$. 
Different curves correspond to different dependence of total power of initial fluctuations on the radial coordinate in the disc: 
$P(R)\propto R^{-2}$ ({$F_{var}\propto 1/R$;} blue dashed-dotted), 
$P(R)\propto R^{-1}$ ({$F_{var}=$constant;} black solid), 
$P(R)\propto R^{-1/2}$ ({$F_{var}\propto R^{1/2}$;} red dashed).
Parameters: $f_{\rm br}=100R^{-3/2}$, $n=0.75$.
}
\label{pic:GreenLipEx07}
\end{figure}

\subsection{Comparison with data}
\label{sec:CompToData}

We compare the theoretical PDS of X-ray flux variability with the observed PDS in two luminosity states {of the} X-ray transient A~0535+26 
(see Fig.\,\ref{pic:DataEx}),
\footnote{{
The data for the plot are from \textit{RXTE}/PCA, which observed A~0535+262 during its 2009 outburst. From each observation, light curves were extracted using {the} {\sc heasoft} software package {\sc XRONOS}. Power spectra were calculated from light curve segments of 512 s duration using the {\sc powspec} tool with Miyamoto normalization and averaged together from observations within a luminosity interval. Poisson noise level was subtracted from the power spectra, which were then multiplied by frequency. The power spectra presented here correspond to observations with average luminosities of $1.7\times10^{35}\,\ergs$ and $3.8\times10^{36}\,\ergs$.}}
which is one of the brightest XRPs on {the} X-ray sky located at distance around 2 kpc from the Sun \citep{1998MNRAS.297L...5S}.
A cyclotron line scattering feature at $E_{\rm cyc,0}\sim 45\,{\rm keV}$ and its first harmonic at $E_{\rm cyc,1}\sim 100\,{\rm keV}$ {have been detected} \citep{2007A&A...465L..21C}.
Thus, A~0535+26 is a source {for which} we can estimate magnetic field strength at the NS surface: $B\sim 4\times 10^{12}\,{\rm G}$ and, therefore, {the} inner disc radius at any given mass accretion rate (see equation \ref{eq:Rm}). 
%\green{AI: Do we not need to know the distance to be able to constrain the intrinsic accretion luminosity?}

\begin{figure}
\centering 
\includegraphics[width=9.cm]{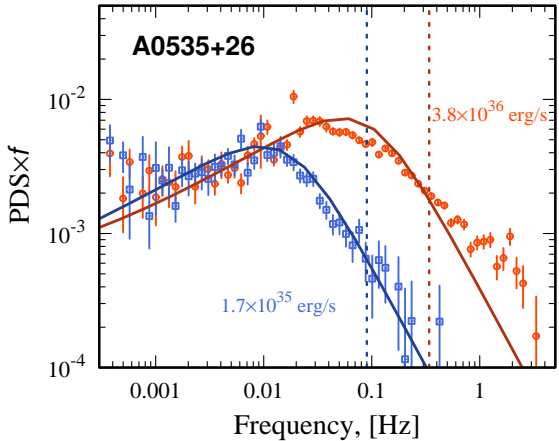}
\caption{{The PDS of X-ray energy flux fluctuations in two luminosity states of {the} X-ray transient A~0535+26 during its outburst in 2009:
$L_1\simeq 1.7\times 10^{35}\,\ergs$ (blue squares) and $L_2 \simeq 3.8\times 10^{36}\,\ergs$ (red circles).
The power spectra exhibit a break, whose frequency varies with the accretion luminosity: the higher the luminosity, the higher the break frequency. 
Blue and red solid lines represent the theoretical PDS calculated on the {basis} of our model of propagating fluctuations of the mass accretion rate in {the} disc.
Blue and red dashed lines represent Keplerian frequencies at the inner disc radii at  $L_1\simeq 1.7\times 10^{35}\,\ergs$ and $L_2 \simeq 3.8\times 10^{36}\,\ergs$ respectively.
In both cases the Keplerian frequency is a few times higher than the frequency of initial fluctuations assumed in the model. 
} 
}
\label{pic:DataEx}
\end{figure}

Using our model, we calculate {the} theoretical PDS of mass accretion rate fluctuations at the inner disc radius and compare it with the observed PDS of X-ray flux variability in two luminosity states of A~0535+26. The input parameters of our model are $n$, $R_{\rm m}$ [in cm], $R_{\rm out}$ [in cm], $f_{\rm br}(R_{\rm m})$ [in Hz], $f_{\rm br}(R_{\rm m})/f_{\rm v}(R_{\rm m})$, $f_0$ and $F_{\rm var}$. We use $n=3/4$ (typical for a gas pressure dominated $\alpha$-disc) and calculate the magnetospheric radius of the two luminosity states by combining the B-field measurement with the observed luminosities of the two states
\beq 
L_1\simeq 1.7\times 10^{35}\,\ergs, \nonumber \\
L_2\simeq 3.8\times 10^{36}\,\ergs. \nonumber
\eeq
To obtain theoretical PDSs we choose the inner disc radii to be
\beq 
R_{\rm m,1}=8.8\times 10^8\,{\rm cm}, \nonumber \\
R_{\rm m,2}=3.6\times 10^8\,{\rm cm}, \nonumber
\eeq 
according to equation (\ref{eq:Rm}). We fix the outher disc radius at $R_{\rm out} = 10^{10}\,{\rm cm}$, but note that the model in the frequency range explored is almost insensitive to this parameter. We set the the break frequencies at the inner radius to
\beq
f_{\rm br,1}=0.01\,{\rm Hz}, \nonumber \\
f_{\rm br,2}=0.07\,{\rm Hz}, \nonumber
\eeq 
and $f_{\rm br}/f_{\rm v}=100$ at the inner disc radius (see Fig.\,\ref{pic:DataEx}).
We fix $f_0=0$ for both observations and set $F_{\rm var}$, which acts simply as a normalisation constant of the PDS, separately for the two observations.

For the above values of $R_{\rm m}$, and assuming $M_{1.4}=1$, the Keplerian frequencies at the inner disc radii are expected to be
\beq
f_{\rm K,1}(R_{\rm m})\approx 0.09\,{\rm Hz}, \nonumber \\
f_{\rm K,2}(R_{\rm m})\approx 0.34\,{\rm Hz}. \nonumber
\eeq
In both cases this is a few times higher than the break frequency used in our modeling (see the gaps between the breaks and Keplerian frequencies in Fig.\,\ref{pic:DataEx}), giving $k_d \sim 9$ and $\sim 5$ for the low and high flux state respectively.
The possibility of an underestimate of the inner disc radius (and corresponding overestimation of the Keplerian frequency) {from} relation (\ref{eq:Rm}) due to uncertainty {in} $\Lambda$ can be largely excluded because the inner disc radius in A~0535+26 was verified by measurements of NS spin evolution during a few outburst in 2009-2015 \citep{2017PASJ...69..100S}, which have shown that the magnetic field is likely dominated by its dipole component and the inner disc radius is well described by the relation (\ref{eq:Rm}) with $\Lambda=0.5$. 
The estimations of the inner radius show that the disc in these particular cases is geometrically thin and gas pressure dominated in the considered time intervals (see Fig.\,\ref{pic:DiscZones}).
Therefore, the exponent $n$ in (\ref{eq:nu(R)}) can be taken to be $3/4$.
}
Thus, we have to conclude that the initial variability is generated at frequencies which are noticeably lower than the local Keplerian frequency. 
The inference that initial fluctuations generated at frequencies {a} few times below the local Keplerian frequency is consistent with the idea that initial fluctuations are driven by {a} dynamo process \citep{2004MNRAS.348..111K} resulting from the magneto-rotational instability \citep{1991ApJ...376..214B}. Thus, the modeling of XRP PDSs provides an opportunity to measure the time scales of the dynamo process in accretion discs observationally. 
However, these measurements require much more detailed analysis of PDS variability with X-ray luminosity, which is beyond the scope of this paper.

\section{Summary and discussion}
\label{sec:Summary}

We have considered {the} physical processes responsible for the broadband aperiodic variability of X-ray energy flux in accreating highly magnetized NSs - XRPs.
In the case of XRPs, emission in {the} X-ray energy band originates from small regions located close to the surface {of the} NS, while the accretion disc is truncated at {a} large distance by a strong magnetic field and {thus} does not produce an appreciable fraction of X-ray energy flux.
\footnote{With the exception of a narrow energy band, where the accretion disc produces {an} iron K$\alpha$ line due to the reflection of X-ray flux from the accretion disc \citep{1991MNRAS.249..352G}.
Note that, the light crossing time of the inner radius is much shorter than the Keplerian time scale, and the variability of the reflected emission should replicate the variability of the accretion rate at the NS surface. 
} 
The observed aperiodic variability of X-ray flux in XRPs is caused by mass accretion rate variability in the vicinity of {the} NS surface, which replicates the variability of the mass accretion rate at the inner radius of the accretion disc. 
Because the majority of X-ray photons originate from a small region at the NS surface, the timing properties of aperiodic variability in X-rays are expected to be independent {of} the energy band. 
Possible dependencies of the timing properties on the photon energy are expected to be caused by further reprocessing of X-ray photons, which beyond the scope of this paper. 
Fluctuations of the mass accretion rate at the inner disc radius are shaped by fluctuations arising all over the disc and propagating inwards due to the process of viscous diffusion \citep{1997MNRAS.292..679L,2001MNRAS.321..759C,2001MNRAS.327..799K}. 
The process of viscous diffusion modifies {the} timing properties of propagating fluctuations, suppressing the variability at high frequencies \citep{2018MNRAS.474.2259M}. 
The diffusion process is determined by properties of viscosity in {the} accretion disc and boundary conditions: displacement of {the} inner and outer radii of the disc and conditions there (local torques and mass accretion rates).

We have developed a theoretical base for calculations of mass accretion rate variability at the inner radius of the accretion disc, which largely shapes the timing properties of XRPs. 
Using known analytical solutions of the equation of viscous diffusion (\ref{eq:DifEqGen}), we have investigated the transfer properties of the discs. 
A new Green function of the viscous diffusion equation accounting for fixed inner and outer radii of {the} accretion disc has been derived (see Section\,\ref{sec:GfLipunova}).
Our Green function generalizes the previous solutions obtained for accretion discs in the Newtonian approximation \citep{1974MNRAS.168..603L,2011MNRAS.410.1007T,2015ApJ...804...87L}.
The obtained Green function provides a more complete description of the propagation of mass accretion rate fluctuations under {the} influence of viscous diffusion in geometrically thin accretion discs.

XRPs are unique objects because they provide {the} possibility to probe timing properties of mass accretion rate variability within a small range of radial coordinates in {the} accretion disc (in contrast to accreting BHs, where the observer detects X-ray photons originating from the extended inner region of the disc, see e.g. \citealt{2013MNRAS.434.1476I,2018MNRAS.474.2259M}). 
The geometry of {the} accretion disc (particularly, the inner disc radius) depends on the mass accretion rate.
{The timing properties of aperiodic variability, and particularly the PDS of the variability, depend on accretion disc geometry and, therefore, on the mass accretion rate.
As a result, transient XRPs and analyses of their variability can be used to probe {the} geometry (its inner radius) of a disc and {the} physical conditions there: {the} time scale of the dynamo process, which is assumed to be responsible for the initial fluctuation of viscosity.}

In the case of known strength and structure of NS magnetic field and, therefore, known inner radius of {the} disc at given luminosity (see Section \ref{sec:ADGeometry}), we can test physical conditions in the accretion disc. 
Particularly, one can measure the time scale of initial fluctuations of the mass accretion rate.
Using the observed PDS of {the} transient X-ray pulsar A~0535+26 and modeling it with our theory, we conclude that the typical frequency of initial fluctuations is lower (by a factor of {$\sim 5-9$}) than the local Keplerian frequency (see Section \ref{sec:CompToData}).
Note, that this conclusion directly follows from the assumption that the break in the PDS corresponds to the characteristic time scale at the inner boundary of the disk, which is not suppressed by the process of viscous diffusion.
This statement is in agreement with the model, where the initial fluctuations are caused by the dynamo process, driven by the magneto-rotational instability in the accretion disc \citep{2004MNRAS.348..111K}. 
The coefficient of proportionality between the Keplerian time scale and dynamo time scale is a matter of first-principal numerical simulations. 
According to our results, this coefficient can be obtained from observations of X-ray transients with known magnetic fields.
This, however, requires detailed analysis of PDS variability with mass accretion rate and is beyond the scope of the paper.

It is worth {noting} that there are still a few open issues in the problem: (i) the exact timing properties of initial fluctuations of the mass accretion rate, and (ii) the timing properties of instabilities developing at the inner disc radius. 
These problems have to be addressed {by} numerical MHD simulations.

%%%%%%%%%%%%%%%%%%%%%%%%%%%%%%%%%%%%%%%%%%%%%%%%%%%%%%%%%%%%%%%%%%%%%%%%%%%%%%
%% Acknowledgments                                                         %%
%%%%%%%%%%%%%%%%%%%%%%%%%%%%%%%%%%%%%%%%%%%%%%%%%%%%%%%%%%%%%%%%%%%%%%%%%%%%%%
\section*{Acknowledgements}

This research was supported by 
the Netherlands Organization for Scientific Research (AAM),
the grant 14.W03.31.0021 of the Ministry of Education and Science of the Russian Federation (AAM and SST), 
RFBR grant 18-502-12025 (GVL),
the Royal Society (AI),
the V\"ais\"al\"a Foundation (SST). 
The authors would like to acknowledge networking support by the COST Actions CA16214 and CA16104.
We are also grateful to Victor Doroshenko and Nikolai Shakura for discussion and a number of useful comments.

%%%%%%%%%%%%%%%%%%%%%%%%%%%%%%%%%%%%%%%%%%%%%%%%%%%%%%%%%%%%%%%%%%%%%%%%%%%%%%
%% Bibliography                                                             %%
%%%%%%%%%%%%%%%%%%%%%%%%%%%%%%%%%%%%%%%%%%%%%%%%%%%%%%%%%%%%%%%%%%%%%%%%%%%%%%
%\bibliographystyle{mn2e}
%\bibliography{allbib}

\begin{thebibliography}{}

\bibitem[\protect\citeauthoryear{{Aly}}{{Aly}}{1980}]{1980A&A....86..192A}
{Aly} J.~J.,  1980, \aap, 86, 192

\bibitem[\protect\citeauthoryear{{Ar{\'e}valo} \& {Uttley}}{{Ar{\'e}valo} \&
  {Uttley}}{2006}]{2006MNRAS.367..801A}
{Ar{\'e}valo} P.,  {Uttley} P.,  2006, \mnras, 367, 801

\bibitem[\protect\citeauthoryear{{Arons}}{{Arons}}{1992}]{1992ApJ...388..561A}
{Arons} J.,  1992, \apj, 388, 561

\bibitem[\protect\citeauthoryear{{Bachetti}}{{Bachetti et al.}}{2014}]{2014Natur.514..202B}
{Bachetti} M. et al.,  2014, Nature, 514, 202

\bibitem[\protect\citeauthoryear{{Balbus}}{{Balbus}}{2017}]{2017MNRAS.471.4832B}
{Balbus} S.~A.,  2017, \mnras, 471, 4832

\bibitem[\protect\citeauthoryear{{Balbus} \& {Hawley}}{{Balbus} \&
  {Hawley}}{1991}]{1991ApJ...376..214B}
{Balbus} S.~A.,  {Hawley} J.~F.,  1991, \apj, 376, 214

\bibitem[\protect\citeauthoryear{{Basko} \& {Sunyaev}}{{Basko} \&
  {Sunyaev}}{1976}]{1976MNRAS.175..395B}
{Basko} M.~M.,  {Sunyaev} R.~A.,  1976, \mnras, 175, 395

\bibitem[\protect\citeauthoryear{{Begelman}}{{Begelman}}{2006}]{2006ApJ...643.1065B}
{Begelman} M.~C.,  2006, \apj, 643, 1065

\bibitem[\protect\citeauthoryear{{Brandenburg}, {Nordlund}, {Stein} \&
  {Torkelsson}}{{Brandenburg} et~al.}{1995}]{1995ApJ...446..741B}
{Brandenburg} A.,  {Nordlund} A.,  {Stein} R.~F.,    {Torkelsson} U.,  1995,
  \apj, 446, 741

\bibitem[\protect\citeauthoryear{{Caballero}}{{Caballero}}{2007}]{2007A&A...465L..21C}
{Caballero} I.,  2007, \aap, 465, L21

\bibitem[\protect\citeauthoryear{{Chashkina}, {Abolmasov} \&
  {Poutanen}}{{Chashkina} et~al.}{2017}]{2017MNRAS.470.2799C}
{Chashkina} A.,  {Abolmasov} P.,    {Poutanen} J.,  2017, \mnras, 470, 2799

\bibitem[\protect\citeauthoryear{{Chashkina}, {Lipunova}, {Abolmasov} \&
  {Poutanen}}{{Chashkina} et~al.}{2019}]{2019arXiv190204609C}
{Chashkina} A.,  {Lipunova} G.,  {Abolmasov} P.,    {Poutanen} J.,  2019, arXiv
  e-prints

\bibitem[\protect\citeauthoryear{{Churazov}, {Gilfanov} \&
  {Revnivtsev}}{{Churazov} et~al.}{2001}]{2001MNRAS.321..759C}
{Churazov} E.,  {Gilfanov} M.,    {Revnivtsev} M.,  2001, \mnras, 321, 759

\bibitem[\protect\citeauthoryear{{Davidson} \& {Ostriker}}{{Davidson} \&
  {Ostriker}}{1973}]{1973ApJ...179..585D}
{Davidson} K.,  {Ostriker} J.~P.,  1973, \apj, 179, 585

\bibitem[\protect\citeauthoryear{{Frank}, {King} \& {Raine}}{{Frank}
  et~al.}{2002}]{2002apa..book.....F}
{Frank} J.,  {King} A.,    {Raine} D.~J.,  2002, {Accretion Power in
  Astrophysics: Third Edition}

\bibitem[\protect\citeauthoryear{{George} \& {Fabian}}{{George} \&
  {Fabian}}{1991}]{1991MNRAS.249..352G}
{George} I.~M.,  {Fabian} A.~C.,  1991, \mnras, 249, 352

\bibitem[\protect\citeauthoryear{{Ghosh} \& {Lamb}}{{Ghosh} \&
  {Lamb}}{1979}]{1979ApJ...232..259G}
{Ghosh} P.,  {Lamb} F.~K.,  1979, \apj, 232, 259

\bibitem[\protect\citeauthoryear{{Giangrande}, {Giovannelli}, {Bartolini},
  {Guarnieri} \& {Piccioni}}{{Giangrande} et~al.}{1980}]{1980A&AS...40..289G}
{Giangrande} A.,  {Giovannelli} F.,  {Bartolini} C.,  {Guarnieri} A.,
  {Piccioni} A.,  1980, \aaps, 40, 289

\bibitem[\protect\citeauthoryear{{Gilfanov} \& {Arefiev}}{{Gilfanov} \&
  {Arefiev}}{2005}]{2005astro.ph..1215G}
{Gilfanov} M.,  {Arefiev} V.,  2005, ArXiv Astrophysics e-prints

\bibitem[\protect\citeauthoryear{{Hawley}, {Gammie} \& {Balbus}}{{Hawley}
  et~al.}{1995}]{1995ApJ...440..742H}
{Hawley} J.~F.,  {Gammie} C.~F.,    {Balbus} S.~A.,  1995, \apj, 440, 742

\bibitem[\protect\citeauthoryear{{Hogg} \& {Reynolds}}{{Hogg} \&
  {Reynolds}}{2016}]{2016ApJ...826...40H}
{Hogg} J.~D.,  {Reynolds} C.~S.,  2016, \apj, 826, 40

\bibitem[\protect\citeauthoryear{{Hoshino} \& {Takeshima}}{{Hoshino} \&
  {Takeshima}}{1993}]{1993ApJ...411L..79H}
{Hoshino} M.,  {Takeshima} T.,  1993, \apjl, 411, L79

\bibitem[\protect\citeauthoryear{{Illarionov} \& {Sunyaev}}{{Illarionov} \&
  {Sunyaev}}{1975}]{1975A&A....39..185I}
{Illarionov} A.~F.,  {Sunyaev} R.~A.,  1975, \aap, 39, 185

\bibitem[\protect\citeauthoryear{{Ingram} \& {Done}}{{Ingram} \&
  {Done}}{2011}]{2011MNRAS.415.2323I}
{Ingram} A.,  {Done} C.,  2011, \mnras, 415, 2323

\bibitem[\protect\citeauthoryear{{Ingram} \& {Done}}{{Ingram} \&
  {Done}}{2012}]{2012MNRAS.419.2369I}
{Ingram} A.,  {Done} C.,  2012, \mnras, 419, 2369

\bibitem[\protect\citeauthoryear{{Ingram} \& {van der Klis}}{{Ingram} \& {van
  der Klis}}{2013}]{2013MNRAS.434.1476I}
{Ingram} A.,  {van der Klis} M.,  2013, \mnras, 434, 1476

\bibitem[\protect\citeauthoryear{{Ingram}}{{Ingram}}{2016}]{2016AN....337..385I}
{Ingram} A.~R.,  2016, Astronomische Nachrichten, 337, 385

\bibitem[\protect\citeauthoryear{{Israel}}{{Israel et al.}}{2017}]{2017Sci...355..817I}
{Israel} G.~L. et al., 2017, Science, 355, 817

\bibitem[\protect\citeauthoryear{{Kaminker}, {Fedorenko} \&
  {Tsygan}}{{Kaminker} et~al.}{1976}]{1976SvA....20..436K}
{Kaminker} A.~D.,  {Fedorenko} V.~N.,    {Tsygan} A.~I.,  1976, \sovast, 20,
  436

\bibitem[\protect\citeauthoryear{{King}, {Lasota} \& {Klu{\'z}niak}}{{King}
  et~al.}{2017}]{2017MNRAS.468L..59K}
{King} A.,  {Lasota} J.-P.,    {Klu{\'z}niak} W.,  2017, \mnras, 468, L59

\bibitem[\protect\citeauthoryear{{King}, {Pringle}, {West} \& {Livio}}{{King}
  et~al.}{2004}]{2004MNRAS.348..111K}
{King} A.~R.,  {Pringle} J.~E.,  {West} R.~G.,    {Livio} M.,  2004, \mnras,
  348, 111

\bibitem[\protect\citeauthoryear{{Kotov}, {Churazov} \& {Gilfanov}}{{Kotov}
  et~al.}{2001}]{2001MNRAS.327..799K}
{Kotov} O.,  {Churazov} E.,    {Gilfanov} M.,  2001, \mnras, 327, 799

\bibitem[\protect\citeauthoryear{{Lai}}{{Lai}}{2014}]{2014EPJWC..6401001L}
{Lai} D.,  2014, in European Physical Journal Web of Conferences Vol.~64 of
  European Physical Journal Web of Conferences, {Theory of Disk Accretion onto
  Magnetic Stars}.
p. 01001

\bibitem[\protect\citeauthoryear{{Lasota}}{{Lasota}}{2001}]{2001NewAR..45..449L}
{Lasota} J.-P.,  2001, \nar, 45, 449

\bibitem[\protect\citeauthoryear{{Lipunov}}{{Lipunov}}{1978}]{1978SvA....22..702L}
{Lipunov} V.~M.,  1978, \sovast, 22, 702

\bibitem[\protect\citeauthoryear{{Lipunov}}{{Lipunov}}{1987}]{1987ans..book.....L}
{Lipunov} V.~M.,  1987, {The astrophysics of neutron stars}

\bibitem[\protect\citeauthoryear{{Lipunova}}{{Lipunova}}{1999}]{1999AstL...25..508L}
{Lipunova} G.~V.,  1999, Astronomy Letters, 25, 508

\bibitem[\protect\citeauthoryear{{Lipunova}}{{Lipunova}}{2015}]{2015ApJ...804...87L}
{Lipunova} G.~V.,  2015, \apj, 804, 87

\bibitem[\protect\citeauthoryear{{Lipunova} \& {Malanchev}}{{Lipunova} \&
  {Malanchev}}{2017}]{2017MNRAS.468.4735L}
{Lipunova} G.~V.,  {Malanchev} K.~L.,  2017, \mnras, 468, 4735

\bibitem[\protect\citeauthoryear{{Liska}, {Tchekhovskoy} \& {Quataert}}{{Liska}
  et~al.}{2018}]{2018arXiv180904608L}
{Liska} M.~T.~P.,  {Tchekhovskoy} A.,    {Quataert} E.,  2018, ArXiv e-prints

\bibitem[\protect\citeauthoryear{{Lynden-Bell} \& {Pringle}}{{Lynden-Bell} \&
  {Pringle}}{1974}]{1974MNRAS.168..603L}
{Lynden-Bell} D.,  {Pringle} J.~E.,  1974, \mnras, 168, 603

\bibitem[\protect\citeauthoryear{{Lyubarskii}}{{Lyubarskii}}{1997}]{1997MNRAS.292..679L}
{Lyubarskii} Y.~E.,  1997, \mnras, 292, 679

\bibitem[\protect\citeauthoryear{{Lyubarskii} \& {Syunyaev}}{{Lyubarskii} \&
  {Syunyaev}}{1988}]{1988SvAL...14..390L}
{Lyubarskii} Y.~E.,  {Syunyaev} R.~A.,  1988, Soviet Astronomy Letters, 14, 390

\bibitem[\protect\citeauthoryear{{McHardy}, {Papadakis}, {Uttley}, {Page} \&
  {Mason}}{{McHardy} et~al.}{2004}]{2004MNRAS.348..783M}
{McHardy} I.~M.,  {Papadakis} I.~E.,  {Uttley} P.,  {Page} M.~J.,    {Mason}
  K.~O.,  2004, \mnras, 348, 783

\bibitem[\protect\citeauthoryear{{Mushtukov}, {Ingram}, {Middleton}, {Nagirner}
  \& {van der Klis}}{{Mushtukov} et~al.}{2019}]{2019MNRAS.484..687M}
{Mushtukov} A.~A.,  {Ingram} A.,  {Middleton} M.,  {Nagirner} D.~I.,    {van
  der Klis} M.,  2019, \mnras, 484, 687

\bibitem[\protect\citeauthoryear{{Mushtukov}, {Ingram} \& {van der
  Klis}}{{Mushtukov} et~al.}{2018}]{2018MNRAS.474.2259M}
{Mushtukov} A.~A.,  {Ingram} A.,    {van der Klis} M.,  2018, \mnras, 474, 2259

\bibitem[\protect\citeauthoryear{{Mushtukov}, {Suleimanov}, {Tsygankov} \&
  {Ingram}}{{Mushtukov} et~al.}{2017}]{2017MNRAS.467.1202M}
{Mushtukov} A.~A.,  {Suleimanov} V.~F.,  {Tsygankov} S.~S.,    {Ingram} A.,
  2017, \mnras, 467, 1202

\bibitem[\protect\citeauthoryear{{Mushtukov}, {Suleimanov}, {Tsygankov} \&
  {Poutanen}}{{Mushtukov} et~al.}{2015a}]{2015MNRAS.454.2539M}
{Mushtukov} A.~A.,  {Suleimanov} V.~F.,  {Tsygankov} S.~S.,    {Poutanen} J.,
  2015a, \mnras, 454, 2539

\bibitem[\protect\citeauthoryear{{Mushtukov}, {Suleimanov}, {Tsygankov} \&
  {Poutanen}}{{Mushtukov} et~al.}{2015b}]{2015MNRAS.447.1847M}
{Mushtukov} A.~A.,  {Suleimanov} V.~F.,  {Tsygankov} S.~S.,    {Poutanen} J.,
  2015b, \mnras, 447, 1847

\bibitem[\protect\citeauthoryear{{Mushtukov}, {Verhagen}, {Tsygankov}, {van der
  Klis}, {Lutovinov} \& {Larchenkova}}{{Mushtukov}
  et~al.}{2018}]{2018MNRAS.474.5425M}
{Mushtukov} A.~A.,  {Verhagen} P.~A.,  {Tsygankov} S.~S.,  {van der Klis} M.,
  {Lutovinov} A.~A.,    {Larchenkova} T.~I.,  2018, \mnras, 474, 5425

\bibitem[\protect\citeauthoryear{{Narayan} \& {Yi}}{{Narayan} \&
  {Yi}}{1995}]{1995ApJ...452..710N}
{Narayan} R.,  {Yi} I.,  1995, \apj, 452, 710

\bibitem[\protect\citeauthoryear{{Paczynski}}{{Paczynski}}{1977}]{1977ApJ...216..822P}
{Paczynski} B.,  1977, \apj, 216, 822

\bibitem[\protect\citeauthoryear{{Paczynski}}{{Paczynski}}{1992}]{1992AcA....42..145P}
{Paczynski} B.,  1992, \actaa, 42, 145

\bibitem[\protect\citeauthoryear{{Papaloizou} \& {Pringle}}{{Papaloizou} \&
  {Pringle}}{1977}]{1977MNRAS.181..441P}
{Papaloizou} J.,  {Pringle} J.~E.,  1977, \mnras, 181, 441

\bibitem[\protect\citeauthoryear{{Poutanen}, {Lipunova}, {Fabrika}, {Butkevich}
  \& {Abolmasov}}{{Poutanen} et~al.}{2007}]{2007MNRAS.377.1187P}
{Poutanen} J.,  {Lipunova} G.,  {Fabrika} S.,  {Butkevich} A.~G.,
  {Abolmasov} P.,  2007, \mnras, 377, 1187

\bibitem[\protect\citeauthoryear{{Poutanen}, {Mushtukov}, {Suleimanov},
  {Tsygankov}, {Nagirner}, {Doroshenko} \& {Lutovinov}}{{Poutanen}
  et~al.}{2013}]{2013ApJ...777..115P}
{Poutanen} J.,  {Mushtukov} A.~A.,  {Suleimanov} V.~F.,  {Tsygankov} S.~S.,
  {Nagirner} D.~I.,  {Doroshenko} V.,    {Lutovinov} A.~A.,  2013, \apj, 777,
  115

\bibitem[\protect\citeauthoryear{{Pringle} \& {Rees}}{{Pringle} \&
  {Rees}}{1972}]{1972A&A....21....1P}
{Pringle} J.~E.,  {Rees} M.~J.,  1972, \aap, 21, 1

\bibitem[\protect\citeauthoryear{{Psaltis} \& {Chakrabarty}}{{Psaltis} \&
  {Chakrabarty}}{1999}]{1999ApJ...521..332P}
{Psaltis} D.,  {Chakrabarty} D.,  1999, \apj, 521, 332

\bibitem[\protect\citeauthoryear{{Qiao} \& {Liu}}{{Qiao} \&
  {Liu}}{2010}]{2010PASJ...62..661Q}
{Qiao} E.,  {Liu} B.~F.,  2010, \pasj, 62, 661

\bibitem[\protect\citeauthoryear{{Reig}}{{Reig}}{2011}]{2011Ap&SS.332....1R}
{Reig} P.,  2011, \apss, 332, 1

\bibitem[\protect\citeauthoryear{{Revnivtsev}, {Churazov}, {Postnov} \&
  {Tsygankov}}{{Revnivtsev} et~al.}{2009}]{2009A&A...507.1211R}
{Revnivtsev} M.,  {Churazov} E.,  {Postnov} K.,    {Tsygankov} S.,  2009, \aap,
  507, 1211

\bibitem[\protect\citeauthoryear{{Revnivtsev}, {Gilfanov} \&
  {Churazov}}{{Revnivtsev} et~al.}{2000}]{2000A&A...363.1013R}
{Revnivtsev} M.,  {Gilfanov} M.,    {Churazov} E.,  2000, \aap, 363, 1013

\bibitem[\protect\citeauthoryear{{Romanova}, {Ustyugova}, {Koldoba} \&
  {Lovelace}}{{Romanova} et~al.}{2004}]{2004ApJ...616L.151R}
{Romanova} M.~M.,  {Ustyugova} G.~V.,  {Koldoba} A.~V.,    {Lovelace} R.~V.~E.,
   2004, \apjl, 616, L151

\bibitem[\protect\citeauthoryear{{Scharlemann}}{{Scharlemann}}{1978}]{1978ApJ...219..617S}
{Scharlemann} E.~T.,  1978, \apj, 219, 617

\bibitem[\protect\citeauthoryear{{Shakura}}{{Shakura}}{1972}]{1972AZh....49..921S}
{Shakura} N.~I.,  1972, \azh, 49, 921

\bibitem[\protect\citeauthoryear{{Shakura} \& {Sunyaev}}{{Shakura} \&
  {Sunyaev}}{1973}]{1973A&A....24..337S}
{Shakura} N.~I.,  {Sunyaev} R.~A.,  1973, \aap, 24, 337

\bibitem[\protect\citeauthoryear{{Spruit} \& {Taam}}{{Spruit} \&
  {Taam}}{1990}]{1990A&A...229..475S}
{Spruit} H.~C.,  {Taam} R.~E.,  1990, \aap, 229, 475

\bibitem[\protect\citeauthoryear{{Steele}, {Negueruela}, {Coe} \&
  {Roche}}{{Steele} et~al.}{1998}]{1998MNRAS.297L...5S}
{Steele} I.~A.,  {Negueruela} I.,  {Coe} M.~J.,    {Roche} P.,  1998, \mnras,
  297, L5

\bibitem[\protect\citeauthoryear{{Stone}, {Hawley}, {Gammie} \&
  {Balbus}}{{Stone} et~al.}{1996}]{1996ApJ...463..656S}
{Stone} J.~M.,  {Hawley} J.~F.,  {Gammie} C.~F.,    {Balbus} S.~A.,  1996,
  \apj, 463, 656

\bibitem[\protect\citeauthoryear{{Sugizaki}, {Mihara}, {Nakajima} \&
  {Makishima}}{{Sugizaki} et~al.}{2017}]{2017PASJ...69..100S}
{Sugizaki} M.,  {Mihara} T.,  {Nakajima} M.,    {Makishima} K.,  2017, \pasj,
  69, 100

\bibitem[\protect\citeauthoryear{{Suleimanov}, {Lipunova} \&
  {Shakura}}{{Suleimanov} et~al.}{2007}]{2007ARep...51..549S}
{Suleimanov} V.~F.,  {Lipunova} G.~V.,    {Shakura} N.~I.,  2007, Astronomy
  Reports, 51, 549

\bibitem[\protect\citeauthoryear{{Sunyaev} \& {Revnivtsev}}{{Sunyaev} \&
  {Revnivtsev}}{2000}]{2000A&A...358..617S}
{Sunyaev} R.,  {Revnivtsev} M.,  2000, \aap, 358, 617

\bibitem[\protect\citeauthoryear{{Syunyaev} \& {Shakura}}{{Syunyaev} \&
  {Shakura}}{1977}]{1977SvAL....3..138S}
{Syunyaev} R.~A.,  {Shakura} N.~I.,  1977, Soviet Astronomy Letters, 3, 138

\bibitem[\protect\citeauthoryear{{Takeshima}, {Dotani}, {Mitsuda} \&
  {Nagase}}{{Takeshima} et~al.}{1994}]{1994ApJ...436..871T}
{Takeshima} T.,  {Dotani} T.,  {Mitsuda} K.,    {Nagase} F.,  1994, \apj, 436,
  871

\bibitem[\protect\citeauthoryear{{Tanaka}}{{Tanaka}}{2011}]{2011MNRAS.410.1007T}
{Tanaka} T.,  2011, \mnras, 410, 1007

\bibitem[\protect\citeauthoryear{{Titarchuk}, {Shaposhnikov} \&
  {Arefiev}}{{Titarchuk} et~al.}{2007}]{2007ApJ...660..556T}
{Titarchuk} L.,  {Shaposhnikov} N.,    {Arefiev} V.,  2007, \apj, 660, 556

\bibitem[\protect\citeauthoryear{{Tout} \& {Pringle}}{{Tout} \&
  {Pringle}}{1992}]{1992MNRAS.259..604T}
{Tout} C.~A.,  {Pringle} J.~E.,  1992, \mnras, 259, 604

\bibitem[\protect\citeauthoryear{{Tsygankov}, {Mushtukov}, {Suleimanov} \&
  {Poutanen}}{{Tsygankov} et~al.}{2016a}]{2016MNRAS.457.1101T}
{Tsygankov} S.~S.,  {Mushtukov} A.~A.,  {Suleimanov} V.~F.,    {Poutanen} J.,
  2016a, \mnras, 457, 1101

\bibitem[\protect\citeauthoryear{{Tsygankov}, {Lutovinov}, {Doroshenko},
  {Mushtukov}, {Suleimanov} \& {Poutanen}}{{Tsygankov}
  et~al.}{2016b}]{2016A&A...593A..16T}
{Tsygankov} S.~S.,  {Lutovinov} A.~A.,  {Doroshenko} V.,  {Mushtukov} A.~A.,
  {Suleimanov} V.,    {Poutanen} J.,  2016b, \aap, 593, A16

\bibitem[\protect\citeauthoryear{{Tsygankov}, {Mushtukov}, {Suleimanov},
  {Doroshenko}, {Abolmasov}, {Lutovinov} \& {Poutanen}}{{Tsygankov}
  et~al.}{2017a}]{2017A&A...608A..17T}
{Tsygankov} S.~S.,  {Mushtukov} A.~A.,  {Suleimanov} V.~F.,  {Doroshenko} V.,
  {Abolmasov} P.~K.,  {Lutovinov} A.~A.,    {Poutanen} J.,  2017a, \aap, 608,
  A17

\bibitem[\protect\citeauthoryear{{Tsygankov}, {Wijnands}, {Lutovinov},
  {Degenaar} \& {Poutanen}}{{Tsygankov} et~al.}{2017b}]{2017MNRAS.470..126T}
{Tsygankov} S.~S.,  {Wijnands} R.,  {Lutovinov} A.~A.,  {Degenaar} N.,
  {Poutanen} J.,  2017b, \mnras


\bibitem[\protect\citeauthoryear{{Walter}, {Lutovinov}, {Bozzo} \&
  {Tsygankov}}{{Walter} et~al.}{2015}]{2015A&ARv..23....2W}
{Walter} R.,  {Lutovinov} A.~A.,  {Bozzo} E.,    {Tsygankov} S.~S.,  2015,
  \aapr, 23, 2

\bibitem[\protect\citeauthoryear{{Wang} \& {Frank}}{{Wang} \&
  {Frank}}{1981}]{1981A&A....93..255W}
{Wang} Y.-M.,  {Frank} J.,  1981, \aap, 93, 255

\bibitem[\protect\citeauthoryear{{Wang} \& {Robertson}}{{Wang} \&
  {Robertson}}{1985}]{1985A&A...151..361W}
{Wang} Y.-M.,  {Robertson} J.~A.,  1985, \aap, 151, 361

\bibitem[\protect\citeauthoryear{{Zel'dovich} \& {Shakura}}{{Zel'dovich} \&
  {Shakura}}{1969}]{1969SvA....13..175Z}
{Zel'dovich} Y.~B.,  {Shakura} N.~I.,  1969, \sovast, 13, 175

\end{thebibliography}

%\comment
{

}

\appendix

\section{Accretion disc with a finite inner radius}
\label{sec:GfTanaka}

{
There are a few known analytical solutions of the equation of viscous diffusion. 
In order to describe the viscous evolution of the accretion disc in an XRP, one should account for a non-zero inner radius of a disc.
Green functions for the case of the non-zero inner disc radius and infinite outer radius ($R_{\rm out}=\infty$) have been derived by \cite{2011MNRAS.410.1007T}. }
In the case of zero torque at $R_{\rm in}$ the Green functions of the  surface density are given by
\beq\label{eq:GfTanaka}
& G(R,R',t)=\left(1-\frac{n}{2}\right)R^{-n-1/4}R'^{5/4}R_{\rm in}^{n-2} & \\
&
\times\int\limits_{0}^{\infty}\frac{F_1(l,k,x)F_1(l,k,x')}{F_2(l,k)}
\exp\left[-2\left(1-\frac{n}{2}\right)^2 k^2\frac{t}{t_{{\rm v,\,in}} }\right]k\d k,
& \nonumber
\eeq
where $x=(R/R_{\rm in})^{1-n/2}$, $t_{\rm v,\, in}=\frac{2}{3}R_{\rm in}^2/\nu(R_{\rm in})$ is the local viscous time at the inner disc radius,
$$F_1(l,k,x)=J_l(kx)Y_l(k)-Y_l(kx)J_l(k),$$
$$F_2(l,k)=J_l^2(k)+Y_l^2(k)$$
and $J_{l}(x)$ and $Y_l(x)$ are the Bessel functions of the first and second kind. 
The corresponding Green function for the mass accretion rate $G_{\dot{M}}(R,R',t)$ can be found according to equation (\ref{eq:mdot02}). 
The Green function for the mass accretion rate in the frequency domain takes the following form:
\beq \label{eq:GfTanaka_f}
 \overline{G}_{\dot{M}}(R,R',f)=12\pi t_0\nu_0 \left(1-\frac{n}{2}\right)R^{1/2}R'^{5/4}R_{\rm in}^{-2}&& \nonumber \\
\times\frac{\partial}{\partial R}
\left[
R^{1/4}\int\limits_{0}^{\infty}\d k\,k\frac{F_1(l,k,x)F_1(l,k,x')}{F_2(l,k)(4\pi i f/f_{in}+k^2(n-2)^2)}
\right].&&
\eeq 
{
The Green functions given by (\ref{eq:GfTanaka}) and (\ref{eq:GfTanaka_f}) ignore the effects arising from the disc truncation at larger radii.
}

\section{Green functions for the disc with finite inner and outer radii}
\label{sec:G_F}

The equation of viscous diffusion in an accretion disc (\ref{eq:DifEqGen}) can be rewritten in terms of the viscous torque 
$$F(R,t) =3\pi h\nu(R)\Sigma(R,t)$$  
and the specific angular momentum $h=(GMR)^{1/2}$ as follows:
\beq\label{eq.vis_dif_eq_F_h_Sigma}
\frac{ \partial F }{\partial t} = \frac{3}{4} \nu_{0} h^{2n-2}
(G M)^{2-n}
\frac{\partial^2 F}{\partial h^2}\, .
\eeq
If the kinematic viscosity $\nu$ is a function of radial coordinate only, as it is determined by equation (\ref{eq:nu(R)}), 
the equation (\ref{eq.vis_dif_eq_F_h_Sigma}) is a linear diffusion equation. 
Its solution can be found in terms of Green functions, specified by the boundary conditions in the disc 
\citep{1974MNRAS.168..603L,2011MNRAS.410.1007T,2015ApJ...804...87L}. 

\subsection{Zero torque at the inner disc radius}

\citet{2011MNRAS.410.1007T} found the form of basis functions which should be used to fulfill a boundary condition at a non-zero $R_{\rm in}$ (see his equation \ref{eq:GfTanaka}). 
Extending a particular solution of \citet{2015ApJ...804...87L} to a case of $R_\mathrm{in}\neq 0$, 
we can construct a Green function $G_F(x,x_1,t)$ for the torque $F$, valid for a disc with the finite inner \emph{and} outer radii:
\beq
\label{eq.green_rin_rout1}
&G_F(x,x_1,t) =\\
&= 2 \, x^l\, x_1^{1-l}\,x_{\rm out} ^{-2} \nonumber \\
&\times\sum\limits_i 
\exp\left({-\frac{k_i^2}{8\,l^2}\, \frac{t}{t_{\rm v}}} \right)\, 
\frac{V(k_i\,x_1,k_i\,x_{\rm in})\,V(k_i\,x,k_i\,x_{\rm in})}{V^2(k_i\,x_{\rm out},k_i\,x_{\rm in})},\nonumber
\eeq
where for $0<l<1$ and  $l= 1/(4-2\,n)$ we use the following combination of the Bessel functions:
\begin{equation}
V(u,v) = J_l (u)\, J_{-l} (v) - J_{-l} (u)\, J_l (v)\, ,
 \label{eq.Vi1}
\end{equation}
the viscous time at the outer disc radius $t_\mathrm{v} = 2 {R_{\rm out}^2}/{(3\nu(R_{\rm out}))}\,$,
$x\equiv (R/R_{\rm out})^{(2-n)/2}$,
and $k_i$ are the roots of the transcendent equation  (notice that $h/h_{\rm out} = \xi = x^{2l}$)
\be
 \frac{\partial [x^l \, V(k_i\,x,k_i\,x_{\rm in}) ]}{\partial x^{2l}} \Big|_{x=x_{\rm out}} =0\, .
 \label{eq.k_i0}
\ee
which is equivalent to
\beq  
l V(k_i x_{\rm out},k_i x_{\rm in})+k_i x_{\rm out} \left. \frac{\partial V_l (u, k_i x_{\rm in})}{\partial u} \right|_{u=k_i x_{\rm out}}=0
\eeq
and then to 
\beq\label{eq:ki_App} 
k_i [J_{l-1}(k_i x_{\rm out})J_{-l}(k_i x_{\rm in})-J_{-l-1}(k_i x_{\rm out})J_l(k_i x_{\rm in})]  \nonumber \\
-\frac{2l}{x_{\rm out}}J_{-l}(k_i x_{\rm out})J_l(k_i x_{\rm in})=0.
\eeq
Equation (\ref{eq:ki_App}) has infinite countable number of roots, which can be found numerically. 

The solution of the viscous diffusion equation is given by
\be\label{eq:GF_def}
F(x,t)=\int\limits_{x_{\rm in}}^{x_{\rm out}}\d x'\,G_F(x,x',t)F(x',t=0). 
\ee
Condition \eqref{eq.k_i0} expresses  the homogeneous outer boundary condition on the accretion rate since $\dot M = \partial F/\partial h$.
Notice that, by setting $x_\mathrm{in}=0$, expression \eqref{eq.green_rin_rout1} is reduced to a Green function found in in~\citet{2015ApJ...804...87L}. 

The Green functions for the surface density $G(R,R',t)$ satisfy equation
\be
\Sigma(R,t)=\int\limits_{R_{\rm in}}^{R_{\rm out}}\d R'\,G(R,R',t)\Sigma(R',t=0).  
\ee
Equation (\ref{eq:GF_def}) can be rewritten as
\be 
\Sigma(x,t)=\int\limits_{x_{\rm in}}^{x_{\rm out}}\d x'\,G_F(x,x',t)\frac{h'\nu(x')}{h\nu(x)} \Sigma(x',t=0) 
\ee
or
\beq 
&\Sigma(R,t)=& \nonumber \\
&\int\limits_{R_{\rm in}}^{R_{\rm out}}\d R'\,
\left(1-\frac{n}{2} \right)R^{-n-\frac{1}{2}}R'^{5/4}R_{\rm out}^{\frac{n}{2}-1}G_F(x,x',t)
\Sigma(x',t=0)& \nonumber
\eeq
Thus, the Green functions describing the evolution of the surface density are given by
\beq
&G(R,R',t)=\left(1-\frac{n}{2} \right)R^{-n-\frac{1}{2}}R'^{5/4}R_{\rm out}^{\frac{n}{2}-1}G_F(x,x',t), \nonumber
\eeq
which results in
\beq
&G(R,R',t)=(2-n) R^{-n-1/4} R'^{5/4}R_{\rm out}^{n-2} \\
&\times\sum\limits_{i}
\exp\left[-2\left(1-\frac{n}{2}\right)^2 k_i^2\frac{t}{t_{\rm v}}\right]\frac{V_l(k_i x',k_i x_{\rm in})V_l(k_i x,k_i x_{\rm in})}{V_l^2(k_i x_{\rm out},k_i x_{\rm in})}. \nonumber
\eeq

{
The Green function for the accretion disc of fixed inner and outer radii is expressed by the infinite series (\ref{eq.green_rin_rout1}), where each term contains a root of equations (\ref{eq:ki_App}). 
Numerically, we use a limited number of terms in  (\ref{eq.green_rin_rout1}). 
The smaller the ratio $(R_{\rm out}/R_{\rm in})$, the smaller the number of terms in a series, which are necessary to reach a certain accuracy.
}

\subsection{Zero mass accretion rate at the inner disc radius}

In the case of the zero mass accretion rate at the inner disc radius $R_{\rm in}$ and, therefore, on the NS surface, which is appropriate for the case of so-called "dead discs" \citep{1977SvAL....3..138S} in the propeller state of accretion, the Green function for the torque is modified:
\beq
& G_{F,\dot{M}(R_{\rm in})=0}(x,x_1,t)\\
& = 2\, (1-l)\, x_1^{1-2\,l} / (x_{\rm out}^{2\,l-2} - x_{\rm 
in }^{2\,l-2}) \nonumber \\
& + 2 \, x^l\, x_1^{1-l}\,x_{\rm out} ^{-2} \nonumber \\
& \times\sum\limits_i 
\exp\left({-\frac{k_i^2}{8\,l^2}\, \frac{t}{t_{\rm v}}} \right)\, 
\frac{V_*(k_i\,x_1,k_i\,x_{\rm in})\,V_*(k_i\,x,k_i\,x_{\rm in})}{V_*^2(k_i\,x_{\rm out},k_i\,x_{\rm in})},\nonumber
\eeq
where the function $V_*(u,v)$ is defined as
$$
V_*(u,v) = J_{-l}(u) \, J_{l-1}(v) + J_l(u) \, J_{1-l} (v),
$$
and $k_i$ are roots of transcendent equation  similar to (\ref{eq.k_i0}), but written for the function $V_*(u,v)$.

%\textbf{{Check the accuracy of numerical solution numerically!}}

% Don't change these lines
\bsp	% typesetting comment
\label{lastpage}
\end{document}